\documentclass[fleqn,usenatbib]{mnras}

\usepackage{newtxtext,newtxmath}

\usepackage[T1]{fontenc}

\DeclareRobustCommand{\VAN}[3]{#2}
\let\VANthebibliography\thebibliography
\def\thebibliography{\DeclareRobustCommand{\VAN}[3]{##3}\VANthebibliography}

\usepackage{todonotes}
\usepackage{graphicx}	% Including figure files
\usepackage{amsmath}	% Advanced maths commands
\usepackage{amssymb}	% Extra maths symbols
\title[Primordial binaries and cluster formation]{Implementing Primordial Binaries in Simulations of Star Cluster Formation with a Hybrid MHD and Direct N-Body Method}

\author[Cournoyer-Cloutier et al.]{
Claude Cournoyer-Cloutier$^{1}$ \thanks{E-mail:cournoyc@mcmaster.ca}, Aaron Tran$^{2}$, Sean Lewis$^{3}$, Joshua E. Wall$^{3}$,
\newauthor 
William E. Harris$^{1}$, Mordecai-Mark Mac Low$^{4,2,3,5}$, Stephen L.~W. McMillan$^{2}$,
\newauthor 
Simon Portegies Zwart$^{6}$, and Alison Sills$^{1}$
\\
$^{1}$Department of Physics and Astronomy, McMaster University, 1280 Main Street W, Hamilton, L8S 4L8, Canada\\
$^{2}$ Department of Astronomy, Columbia University, 550 W 120th Street, MC 5246, New York, NY 10027, USA\\
$^{3}$ Department of Physics, Drexel University, Philadelphia, PA 19104, USA\\
$^{4}$ Department of Astrophysics, American Museum of Natural History, New York, NY 10024, USA \\
$^{5}$ Center for Computational Astrophysics, Flatiron Institute, New York, NY 10010, USA\\
$^{6}$ Leiden Observatory, Leiden University, PO Box 9513, NL-2300 RA Leiden, the Netherlands\\
}

\date{Accepted XXX. Received YYY; in original form ZZZ}

\pubyear{2020}

\begin{document}
\label{firstpage}
\pagerange{\pageref{firstpage}--\pageref{lastpage}}
\maketitle

\begin{abstract}
The fraction of stars in binary systems within star clusters is important for their evolution, but what proportion of binaries form by dynamical processes after initial stellar accretion remains unknown. 
In previous work, we showed that dynamical interactions alone produced too few low-mass binaries compared to observations.
We therefore implement an initial population of binaries in the coupled MHD and direct N-body star cluster formation code \texttt{Torch}. 
We compare simulations with, and without, initial binary populations and
follow the dynamical evolution of the binary population in both sets of simulations, finding that both dynamical formation and destruction of binaries take place. Even in the first few million years of star formation, 
we find that an initial population of binaries is needed at all masses to reproduce observed binary fractions for binaries with mass ratios above the $q \geq 0.1$ detection limit. Our simulations also indicate that dynamical interactions in the presence of gas during cluster formation modify the initial distributions towards binaries with smaller primary masses, larger mass ratios, smaller semi-major axes and larger eccentricities. Systems formed dynamically do not have the same properties as the initial systems, and systems formed dynamically in the presence of an initial population of binaries differ from those formed in simulations with single stars only.  Dynamical interactions during the earliest stages of star cluster formation are important for determining the properties of binary star systems.

\end{abstract}

\begin{keywords}
stars: formation -- star clusters: general -- stars: binaries 
\end{keywords}

%%%%%%%%%%%%%%%%%%%%%%%%%%%%%%%%%%%%%%%%%%%%%%%%%%

%%%%%%%%%%%%%%%%% BODY OF PAPER %%%%%%%%%%%%%%%%%%

\section{Introduction}
A complete picture of star cluster formation must account simultaneously for stars forming on the sub-AU scale, stellar dynamics taking place on the cluster's scale and gas flows at the scale of the surrounding giant molecular cloud.  Even when star formation is resolved by a sub-grid model, as is most often the case in simulations, close dynamical encounters between stars must be resolved at the same time as star-gas interactions and large scale stellar dynamics. Effective numerical modelling of cluster formation must therefore be highly multi-scale. Despite these challenges, it is essential to address the problem of star cluster formation, as most stars are formed in a clustered environment~\citep{Lada2003, PortegiesZwart2010}.

Recent reviews of stellar multiplicity in the Galactic field~\citep{Duchene2013, Moe2017} and of protostars embedded in gas~\citep{Reipurth2014} show that most stars, at all evolutionary stages, live in binaries or higher order systems. Surveys of low mass stars~\citep[e.g.][]{Fischer1992, Reid1997, Delfosse2004, winters}, solar-type stars~\citep[e.g.][]{Abt1976, Duquennoy1991, Raghavan2010} and intermediate and high mass stars~\citep[e.g.][]{Sana2011, Sana2012, Chini2012} also reveal a correlation between multiplicity and stellar mass. Both the fraction of stars in multiple systems and the average number of companions per primary increase with increasing primary mass: about 27\% of low mass stars are in multiple systems~\citep{Delfosse2004, winters}, while multiplicity fraction is about 45\% for solar-type~\citep{Raghavan2010} and A-type~\citep{DeRosa2014} stars, and is larger than 90\% for high mass stars~\citep[][and references therein]{Moe2017}.

Despite the ubiquity of binary systems, simulations of star cluster formation and dynamical evolution often use simplistic prescriptions for primordial binaries~\citep[i.e. binaries formed during star formation, e.g.][]{Kroupa1995, Sills1999, PortegiesZwart2001, Leigh2013, Rastello2020} or ignore them altogether~\citep[e.g.][]{PortegiesZwart1999, Pelupessy2012, Sills2018, wall1}, primarily because primordial binaries remain poorly understood via either observations or simulations.
Most observations of binaries in star forming regions~\citep[e.g.][]{Kouwenhoven2005, Reipurth2007, King2012} are of visual binaries, with intermediate separation; binaries with smaller or larger separations are hard to observe. Nonetheless, a significant proportion of stars in star forming regions and in clusters are found in binary systems. Observations of stellar multiplicity in protostars indicate that binary fraction decreases with age, which is attributed to dynamical interactions between the stars~\citep{Tobin2016a}. 

Multiplicity is also influenced by environment.
Binarity in globular clusters is anti-correlated with cluster luminosity~\citep{Milone2016}, and binarity in open clusters is anti-correlated with cluster density~\citep{Duchene1999}.
Young clusters have field-like binary fractions~\citep{Duchene1999, Duchene2018, Sana2011}, and there is no clear difference between the distributions of periods, mass ratios and eccentricities in the field and in young clusters for massive stars~\citep{Sana2011}.
Conversely, loose stellar associations have binary fractions higher than in the field~\citep{Duchene1999, Duchene2018}. The presence of binary systems in star clusters influences their dynamical evolution, for example by facilitating evaporation.
Binaries with low binding energy are disrupted, while energetic binaries become more tightly bound and transfer kinetic energy to the cluster, thus accelerating its dissolution~\citep[e.g.][]{Heggie1975, Hills1975}. 
Appropriate choices of sub-grid model for binary formation and binary parameters -- such as the separation or mass ratio of the generated systems -- are therefore also required for realistic star cluster formation simulations.

The fact that binary systems can be both formed~\citep[e.g.][]{Kouwenhoven2010, Parker2014} and destroyed~\citep[e.g.][]{parker, Parker2012} by the evolution of young clusters further complicates the problem. Although a reasonable assumption would be that some separations (and hence some periods) are associated with primordial formation and others with dynamical formation, it is not so simple. Simulations~\citep[e.g.][]{offner, Sigalotti2018} and observations~\citep[e.g.][]{tobin2016b, Lee2017} show that turbulent core fragmentation and disk fragmentation are viable mechanisms to form binaries during star formation, with separations up to $\sim 1000$ AU. Simulations have also shown that binaries with semi-major axes between 1000 AU and 0.1 pc can be formed during the dissolution of young star clusters~\citep{Kouwenhoven2010}.
\cite{tokovinin} argues that binaries with such separations are more prevalent than what would be predicted by dynamical interactions alone, and proposes that stars forming in adjacent cores could be bound as primordial binaries.
Conversely, dynamical interactions in a young cluster can also form binaries with separations well below 1000 AU~\citep[e.g.][]{Parker2014, wall1}.

We develop a new binary generation algorithm consistent with observations of mass dependent binary fraction and distributions of orbital periods, mass ratios and eccentricities. As an ansatz, we use the observed distribution of zero-age main sequence binary systems in the 
Galactic field to generate our population. Our choice is motivated by the quality of the observations for this population and by the simulations conducted by~\cite{Parker2014}: with pure N-body simulations of star forming regions, they find that using the distributions of binary fraction, mass ratio and period in the field as initial conditions can reproduce the field distribution after dynamical evolution. Our distributions can however be readily modified to investigate different primordial binary distributions. We use the star cluster formation code \texttt{Torch}~\citep{wall1} to demonstrate the impacts of our new binary generation algorithm on the earliest stages of star cluster formation, up to the formation of the first massive stars. 

In Section~\ref{sec:methods}, we describe our simulation environment and our binary generation algorithm. In Section~\ref{sec:simulations}, we present our suite of simulations. In Section~\ref{sec:discussion}, we compare the properties of binary systems in the simulations including primordial binaries and in those starting with only single stars. We summarize our results {in Section~\ref{sec:conclusion} and discuss their implications in Section~\ref{sec:discussion2}.

\section{Methods}\label{sec:methods}

\subsection{Simulating cluster formation with Torch}\label{sec:torch}
~\texttt{Torch}\footnote{\texttt{\url{https://bitbucket.org/torch-sf/torch/branch/binaries} \\ commit 28a27574f667e8a580fe964f5ff185d4fb63f1e7}} uses the \texttt{AMUSE} framework~\citep{amuse} to couple self-gravitating, magnetized gas modelled by the magnetohydrodynamics (MHD) adaptive mesh refinement (AMR) code \texttt{FLASH}~\citep{Fryxell2000} with the N-body code \texttt{ph4}~\citep{McMillan2012} and the stellar evolution code \texttt{SeBa}~\citep{seba}. We use \texttt{FLASH} with a Harten–Lax–van Leer Riemann solver resolving discontinuities~\citep[HLLD,][]{Miyoshi2005} and an unsplit MHD solver~\citep{Lee2013} with third order piecewise parabolic method (PPM) reconstruction~\citep{Colella1984} for gas dynamics, and a multigrid solver for gravity~\citep{Ricker2008}. We handle the gravitational effects of the gas and the stars on one another by a leapfrog integration between the two systems~\citep[see][]{wall1}. Similar gravity bridges have been used previously to couple direct N-body codes with smoothed particle hydrodynamics (SPH) codes~\citep[e.g.][]{Pelupessy2012, Sills2018} and with the AMR code RAMSES~\citep{Gavagnin2017}.

\texttt{Torch} is also optimized to deal with multiple stellar systems. Resolving repeated close encounters between the members of a stable, unperturbed system (e.g. a binary or a hierarchical triple) with the N-body integrator prohibitively shortens the timestep. For each binary or higher order system deemed stable by the Mardling criterion~\citep[][by which triples can have at most one orbital resonance to avoid instability due to large energy exchanges between the orbits]{Mardling2008}, 
we use \texttt{multiples}~\citep{amuse}, which replaces the stars by the systems' centres of mass in \texttt{ph4}. The internal configuration of the system is saved, and the positions of the stars within the system are only computed if the system is perturbed. The encounter between the system and the perturbing star is then resolved with the few-body solver \texttt{smallN}~\citep{Hut1995, McMillan1996}.

Star formation is handled by a sub-grid model via sink particles, which are formed in regions of high local gas density and converging flows, following Jeans' criterion and the additional conditions detailed in~\cite{Federrath2010}. When a sink forms, we use Poisson sampling to generate a list of stars it will form by drawing stellar masses from a~\cite{kroupa} initial mass function~\citep[IMF, ][]{Sormani2017, wall1}, with a minimum sampling mass of 0.08 M$_{\odot}$ and a maximum sampling mass of 150 M$_{\odot}$.
We randomize the list of stars; each star is then formed in order when the sink has accreted sufficient mass. Once it is formed, the sink follows the location of the centre of mass of the local stars and gas, and continues accreting gas. 

\texttt{Torch} also includes stellar feedback, heating and cooling, which are handled via sub-grid models. The amount and location of the feedback depends on the evolution (via \texttt{SeBa}) of the specific stars formed in the simulation. It uses the \texttt{FLASH} module \texttt{FERVENT}~\citep{fervent} for photoionization, direct ultra-violet (UV) radiation pressure from massive stars and photoelectric heating from far-UV radiation. 
It uses the method of \cite{wall2} for stellar winds,
and does not include either indirect radiation pressure or protostellar outflows.

\subsection{Binary generation algorithm}\label{sec:algorithm}
We want to generate a final stellar population that is consistent with the observed IMF, and that also ultimately reproduces the observed binary properties after cluster interactions. However, the effects of the cluster interactions on the primordial binary population are still poorly understood, and observations are not sufficient to have a complete and accurate picture of the properties of binary systems at birth. Nonetheless, observations~\citep[e.g.][]{Sana2011} and simulations~\citep[e.g.][]{Parker2014} suggest that the multiplicity fraction and period, mass ratio and eccentricity distributions in young clusters are consistent with the field population. Volume-limited observations of binary systems in the galactic field, for systems with mass ratios $M_2/M_1 \geq 0.1$, are complete for a very large range of orbital periods~\citep{Moe2017, winters}. 
They are also obtained from much larger samples than observations of young clusters. We therefore adopt for our first suite of simulations a population of primordial binaries with mass-dependent binary fraction and properties consistent with observations of zero-age main sequence stars in binary systems in the Galactic field. Our framework can also be adapted to explore other primordial binary populations. 

\subsubsection{Mass-dependent binary fraction}\label{sec:binary_fraction} 
For simplicity, and following previous studies of binary population synthesis~\citep[e.g.][]{kroupa, Kouwenhoven2009, Parker2014}, we do not form any triple or quadruple systems primordially. 
These are known to be ubiquitous for B and O-type primaries~\citep[e.g.][]{Sana2012, Moe2015}, but represent only 3\% of systems for M-dwarfs~\citep{winters} and 10\% of systems for solar-type stars~\citep{Moe2017}, which account together for $ > 90\%$ of main-sequence stars~\citep{kroupa}.
We treat the mass dependent multiplicity fraction as a mass dependent binary fraction, in order to include all systems included in studies of stellar multiplicity. Since it is hard to determine observationally if there are any unresolved components to a system, most reviews of stellar multiplicity make no distinction among binaries, triples and higher order systems in their distributions of multiplicity fraction, period, mass ratio and eccentricity.
We hence implement a mass dependent binary fraction, which reflects observed distributions of multiplicity fraction. 

For each list of stellar masses obtained at the formation of a sink, we treat each star as a potential primary, and use the primary mass dependent binary fraction to determine if the star is in a binary system. Single stars and primaries are therefore drawn directly from the IMF, while companions are drawn from mass ratio distributions. For each potential primary, we use a random number generator to obtain a number between 0 and 1; the star is found to be in a binary system if the random number is below the mass dependent multiplicity fraction. After a large number of draws, the binary fraction approaches the prescribed multiplicity fraction.

For low-mass stars, we use the observed multiplicity fraction of M-dwarfs in the solar neighbourhood, for primary masses in the mass bins 0.08 -- 0.15~M$_{\odot}$, 0.15 -- 0.30~M$_{\odot}$ and 0.30 -- 0.60~M$_{\odot}$~\citep{winters}. For solar-type stars and above, we use the observed multiplicity fractions for primary masses 0.8 -- 1.2~M$_{\odot}$, 2 -- 5~M$_{\odot}$, 5~--~9~M$_{\odot}$, 9~--~16~M$_{\odot}$ and above 16~M$_{\odot}$~\citep{Moe2017}. Between 0.6~M$_{\odot}$ and 0.8~M$_{\odot}$, and between 1.2~M$_{\odot}$ and 2~M$_{\odot}$, we interpolate linearly between the observed multiplicity fractions. We summarize the multiplicity fractions in Tables~\ref{tab:properties} and~\ref{tab:properties2}.

\begin{table}
	\centering
	\caption{Multiplicity properties from~\citet{winters}. $M_1$ is the primary mass, $\mathcal{F}$ is the binary fraction, $\mu_a$ is the mean projected separation around which the lognormal probability distribution is centered, $\mu_P$ (days) is the corresponding period in days (assuming a circular orbit) and $\log \sigma_P$ is the standard deviation of the lognormal distribution. With the exception of the binary fraction and the period, all the other properties for systems with $M_1 \leq 0.60$ M$_{\odot}$ are obtained from the same distributions as systems with $M_1 \sim 1$ M$_{\odot}$ (see Table~\ref{tab:properties2}, top row).} 
	\label{tab:properties}
	\begin{tabular}{cccccccc}
		$M_1$ & $\mathcal{F}$ & $\mu_a$ (AU) & $\mu_P$ (days) & $\log \sigma_P$ \\
		\hline
		$0.08 - 0.15$ M$_{\odot}$  & 0.16 & 7 & $10^{3.83}$ & $4.12$\\
		$0.15-0.30$ M$_{\odot}$  & 0.21 & 11 & $10^{4.12}$ & $4.37$\\
		$0.30-0.60$ M$_{\odot}$  & 0.28 & 49 & $10^{5.10}$ & $4.78$\\
		\hline
	\end{tabular}
\end{table}

\begin{table}
	\centering
	\caption{Multiplicity properties from~\citet{Moe2017}. $M_1$ is the primary mass, $\mathcal{F}$ is the binary fraction, $P$ is the period range, $\mathcal{F}_P$ is the relative probability for a system to have a period in a given range; $\gamma_{\geq 0.3}$ is the power-law exponent of the mass ratio distribution for $q \geq 0.3$ and $\gamma_{< 0.3}$ is the power-law exponent of the mass ratio distribution for $q < 0.3$.} 
	\label{tab:properties2}
	\begin{tabular}{ccccccc}
		$M_1$ & $\mathcal{F}$ & $P$ (days) & \,\, $\mathcal{F}_P$  \,\, & \,\, $\gamma_{\geq 0.3}$ \,\, & \,\, $\gamma_{< 0.3}$ \,\, \\
		\hline
		$0.8-1.2$ M$_{\odot}$  & 0.40 & $10^{0.5-1.5}$ & 0.06 & $-0.5$ & 0.3\\
		& & $10^{2.5-3.5}$ & 0.13 & $-0.5$ & 0.3\\ 
		& & $10^{4.5-5.5}$ & 0.22 & $-0.5$ & 0.3\\
		& & $10^{6.5-7.5}$ & 0.17 & $-1.1$ & 0.3\\
		\hline
		$2-5$ M$_{\odot}$  & 0.59 & $10^{0.5-1.5}$ & 0.10 & $-0.5$ & 0.2\\
		& & $10^{2.5-3.5}$ & 0.16 & $-0.9$ & 0.1\\ 
		& & $10^{4.5-5.5}$ & 0.18 & $-1.4$ & $-0.5$\\
		& & $10^{6.5-7.5}$ & 0.12 & $-2.0$ & $-1.0$\\
		\hline
		$5-9$ M$_{\odot}$  & 0.76 & $10^{0.5-1.5}$ & 0.11 & $-0.5$ & 0.1\\
		& & $10^{2.5-3.5}$ & 0.18 & $-1.7$ & $-0.2$\\ 
		& & $10^{4.5-5.5}$ & 0.16 & $-2.0$ & $-1.2$\\
		& & $10^{6.5-7.5}$ & 0.09 & $-2.0$ & $-1.5$\\
		\hline
		$9-16$ M$_{\odot}$  & 0.84 & $10^{0.5-1.5}$ & 0.13 & $-0.5$ & 0.1\\
		& & $10^{2.5-3.5}$ & 0.17 & $-1.7$ & $-0.2$\\ 
		& & $10^{4.5-5.5}$ & 0.15 & $-2.0$ & $-1.2$\\
		& & $10^{6.5-7.5}$ & 0.09 & $-2.0$ & $-2.0$\\
		\hline
		$\geq 16$ M$_{\odot}$  & 0.94 & $10^{0.5-1.5}$ & 0.14 & $-0.5$ & 0.1\\
		& & $10^{2.5-3.5}$ & 0.16 & $-1.7$ & $-0.2$\\ 
		& & $10^{4.5-5.5}$ & 0.15 & $-2.0$ & $-1.2$\\
		& & $10^{6.5-7.5}$ & 0.09 & $-2.0$ & $-2.0$\\
		\hline
	\end{tabular}
\end{table}

\subsubsection{Period distribution} 
Periods also depend on primary mass. For primary masses below 0.60~M$_{\odot}$, we use the lognormal distributions from~\cite{winters}, which are given for each of the primary mass bins discussed above. 
For each primary mass range,~\cite{Moe2017} give probabilities at different period values; we extend each given value over one order of magnitude in period (in days), then linearly interpolate between two different period ranges. We use the same mass bins as defined above, but extend the 0.8~--~1.2~M$_{\odot}$ range down to 0.6~M$_{\odot}$ and up to 1.6~M$_{\odot}$, while we extend the 2~--~5~M$_{\odot}$ range down to 1.6~M$_{\odot}$.
We therefore have a probability distribution depending on both primary mass and period.
For each primary, we obtain the orbital period by drawing it from the observed probability distribution for the corresponding mass range, sampled with the rejection method~\citep{Neumann1951}. For each primary, we pick a pair of random numbers -- here, a period between $10^{0.5}$ and $10^{7.5}$ days and a number between 0 and the maximum value of our probability distribution surface -- and accept the pair if the point it defines in parameter space lies below our probability distribution. If it lies above our probability distribution, we reject the pair and repeat the algorithm until a pair is accepted~\citep[following the algorithm from][\S 7.3.6]{recipes}. 

\subsubsection{Companion mass distribution}\label{subsec:mass_ratios}
We obtain the companion masses from distributions of mass ratios $q = M_2/ M_1$, where $M_1$ is the primary mass, $M_2$ is the companion mass and $q \leq 1$ by definition. 
~\cite{Kouwenhoven2009}, in a review of binary pairing functions, summarize as follows the different possible ways to assemble a binary system:
\begin{enumerate}
    \item \textit{Random pairing} \,\,\,\,\,
    Two stars are independently drawn from the IMF; the most massive is labelled as the primary. Random pairing of stars from the Kroupa IMF results in a uniform distribution of system masses~\citep{kroupa}, which~\cite{Kouwenhoven2009} find to result in mass ratios inconsistent with observations. 
    \item \textit{Primary-constrained random pairing} \,\,\,\,\,
    The primary is drawn randomly from the IMF; the companion is then also drawn from the IMF, but with the constraint that it must be less massive than the primary. This pairing function does not reproduce observed mass ratios either~\citep{Kouwenhoven2009}. 
    \item \textit{Primary-constrained pairing} \,\,\,\,\,
    The primary is drawn randomly from the IMF; the companion is then drawn from the mass ratio probability distribution. This technique is meant to be used with a stellar IMF~\citep[e.g.][]{kroupa}. It is compatible with observations, and allows for the use of a primary mass dependent mass ratio distribution, which is observed in nature.
    \item \textit{Split-core pairing} \,\,\,\,\,
    The system mass is drawn randomly from a distribution of system or core masses, then fragmented as a mass ratio is drawn from an observed probability distribution. This technique is meant to be used with a system initial mass function~\citep[e.g.][]{Chabrier2003}. It is also compatible with observations.
\end{enumerate}
Both primary-constrained pairing (iii) and split-core pairing (iv) can reproduce observations of stellar masses and mass ratios concurrently, as well as of a mass dependent binary fraction. They require different pieces of information to implement. Primary-constrained pairing requires a distribution of stellar masses, and primary mass dependent binary properties; split-core pairing requires a distribution of system masses, and can be implemented with primary mass dependent binary properties. We choose to assemble the binary systems with primary-constrained pairing, by drawing primary masses from a~\cite{kroupa} IMF then obtaining the companion masses from the observed primary mass and period-dependent mass ratio distributions. \texttt{Torch} uses by default the~\cite{kroupa} initial mass function, and it is the initial mass function that was used in the original suite of simulations~\citep{wall1, wall2}. We use the same initial mass function for ease of comparison and consistency.

We use the probability distributions from~\cite{Moe2017}, which we extend to lower masses. The mass ratio probability distributions are modelled as power laws,
\begin{equation}
    p_q \propto q^{\gamma}
\end{equation}
where the exponent $\gamma$ is a function of the mass ratio range, the primary mass and the orbital period. We consider three primary mass ranges, $0.08-2$ M$_{\odot}$, $2-5$ M$_{\odot}$ and above 5 M$_{\odot}$; the first mass range is extended from the $0.8-1.2$ M$_{\odot}$ range provided by~\cite{Moe2017} since~\cite{winters} admit that their results are likely incomplete at low companion masses. For each of these mass ranges, we consider a broken power law, with $\gamma$ defined for mass ratios between 0.1 and 0.3, and above 0.3. Finally, the probability is given at different values of the period, between which we interpolate with the same technique as for the period probability distribution. From there, we use the rejection method to obtain a mass ratio. 

We reject mass ratios that would result in substellar companions. We also note that observations are unreliable below $q \leq 0.1$~\citep[][]{Duchene2013, Moe2017, winters}. ~\citet{Price-Whelan2020}, in their analysis of 20,000 close binary systems, acknowledge that their observations are incomplete at low mass ratios. At the high mass end, the problem is most prevalent for spectroscopic searches at intermediate separations~\citep{Kobulnicky2014}. In open clusters, \citet{Sana2011} are only confident in their observations for $q \geq 0.2$ for massive binaries, while~\citet{Deacon2020} are unable to detect companions with $q \leq 0.1$ and estimate that they detect only $\sim 50\%$ of systems with $q = 0.3$ in their surveys of wide binaries in Alpha Per, the Pleiades and Praesepe. Following the completeness limit of~\citet{Moe2017} and~\citet{winters}, we therefore also reject mass ratios below $q = 0.1$.

\subsubsection{Eccentricity distribution}
The eccentricity probability distribution is similarly modelled as a broken power law, as a function of primary mass and period,
\begin{equation}
    p_e \propto e^{\eta}
\end{equation}
where $\eta = 1$ would result in a thermal distribution and $\eta = 0$ would result in a uniform distribution. Following~\cite{Moe2017}, we define 
\begin{equation}\label{eq:eta_low}
    \eta = 0.6 - \frac{0.7}{(\log (P / \text{days}) - 0.5)} 
\end{equation}
for primary masses up to 5 M$_{\odot}$, while for primary masses above 5 M$_{\odot}$, we define
\begin{equation}\label{eq:eta_high}
    \eta = 0.9 - \frac{0.2}{(\log (P / \text{days}) - 0.5)}.
\end{equation}
We further define a period-dependent upper limit on the eccentricity, to avoid binary systems with filled Roche lobes. We use the analytic form of the maximum eccentricity from~\cite{Moe2017},
\begin{equation}
    e_{max}(P) = 1 - \Bigg(\frac{P}{2~\text{days}}\Bigg)^{-2/3}
\end{equation}
which is defined for orbital periods longer than 2 days; we assume all binary systems with shorter periods are circularized~\citep{Raghavan2010}. We use the rejection method to obtain the eccentricities.

\subsubsection{Algorithm test}\label{sec:algorithm_test} 
We test our algorithm by generating a list of stars starting from an initial mass function normalized to 10,000 M$_{\odot}$. We then apply our algorithm to determine which stars are in binary systems; we obtain the period, companion mass and eccentricity for each binary system. We verify that our algorithm indeed reproduces the mass dependent binary fraction observed in the galactic field (Figure~\ref{fig:algorithm_fraction}) and compare our full stellar mass distribution to the initial mass function (Figure~\ref{fig:algorithm_imf}). We also consider our distributions in the primary mass vs. mass ratio parameter space (Figure~\ref{fig:algorithm_mass_ratio}). We generate no systems with a mass ratio $q < 0.1$ and form no stars with a mass below the hydrogen burning limit. 

\begin{figure}
	\includegraphics[width=\columnwidth, clip=true, trim=0.8cm 0 2cm 0]{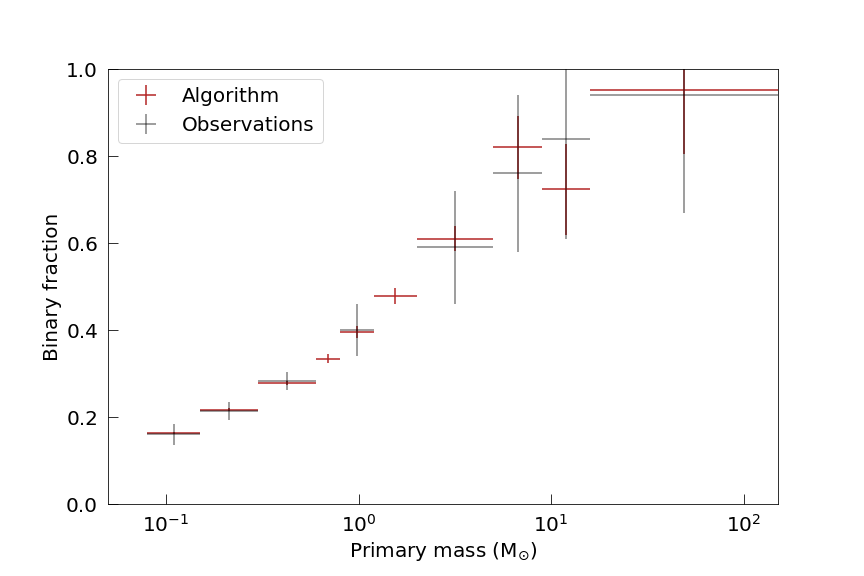}
    \caption{Mass dependent binary fraction, for main sequence stars in the solar neighbourhood~\citep{Moe2017, winters} and for our binary generation algorithm. The errors in $x$ correspond to the bin widths; the errors in $y$ in the observations are from the observational uncertainties while the errors in $y$ on the algorithm data are from the Poisson statistical error.}
    \label{fig:algorithm_fraction}
\end{figure}

\begin{figure}
	\includegraphics[width=\columnwidth, clip=true, trim=0.8cm 0 2cm 0]{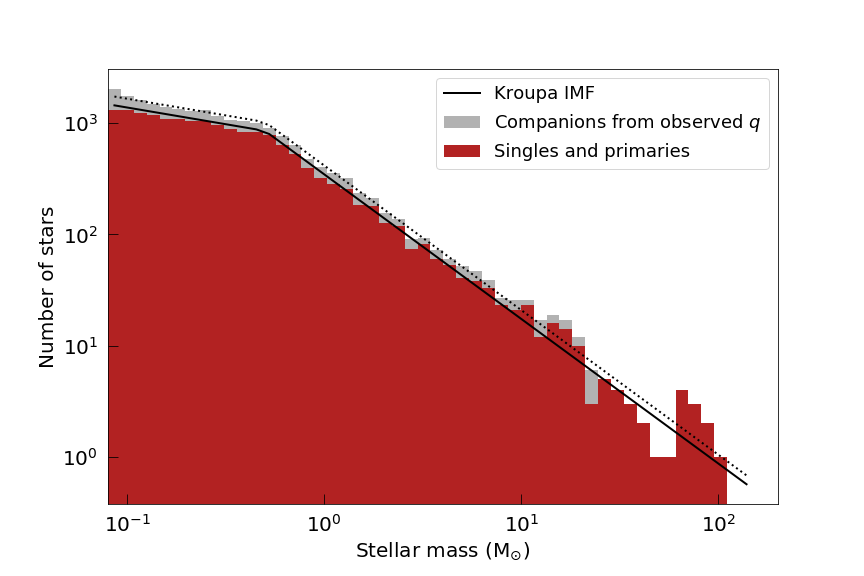}
    \caption{Distribution of stellar masses, for the single stars and primaries drawn from the Kroupa initial mass function (red) and the companions obtained from the mass ratio probability distribution (grey). The Kroupa initial mass function normalized to 10,000 M$_{\odot}$ (solid line) and to the total stellar mass (dotted line) is provided as a guide for the eye.}
    \label{fig:algorithm_imf}
\end{figure}

\begin{figure}
	\includegraphics[width=\columnwidth, clip=true, trim=0.8cm 0 2cm 0]{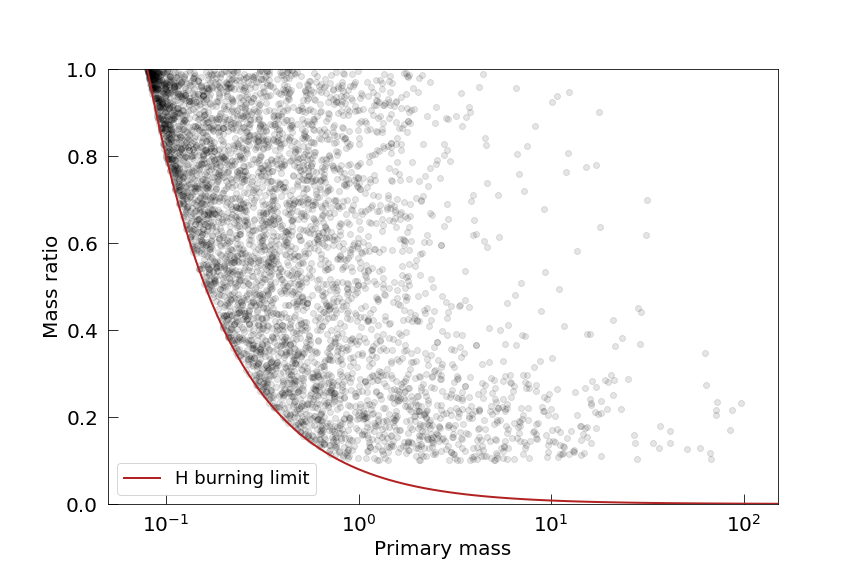}
    \caption{Distribution of mass ratios against primary masses. Our algorithm restricts mass ratios to be higher than 0.1, as observations are highly incomplete below this value. The red line denotes the companion mass corresponding to the hydrogen-burning limit; note that no substellar companions are generated.}
    \label{fig:algorithm_mass_ratio}
\end{figure}

\subsection{Implementing primordial binaries in Torch}
To place a star within the simulation, it must be the next star in the list of stars to be formed by a sink particle, and the sink must have accreted a gas mass equal to or greater than the star's mass. When a star is formed, its mass is subtracted from the sink mass. The star thus formed can be located within the sink's accretion radius, but will be treated by the simulation as a particle distinct from the sink. 
The local gas temperature must be below 100 K at the time of star formation; if the gas temperature is higher, star formation is delayed. Assuming primordial binaries formed through disk or core fragmentation would become stars at the same time, we ensure that stars formed in a binary system are formed at the same time in the simulation. We therefore modify the condition for star formation to require that the mass of a system (whether a single star or a binary) must be accreted by the sink particle before either star is formed. This additional mechanism does not modify the routine to form single stars, but ensures that primaries and their companions are formed simultaneously in the simulation.

The binary systems we generate with our algorithm must be introduced in the simulation with positions and velocities consistent with their orbital properties. We randomize the orbit's orientation by randomizing the inclination, longitude of the ascending node and argument of periapsis, and obtain the relative positions and velocities of the stars by picking a random time in the orbit. The locations of the binaries' centres of mass are chosen in the same way as single stars' positions in \texttt{Torch}~\citep{wall1}.
The position of each star within the simulation domain is finally obtained by adding together the position of the sink in which it forms, the position vector of the centre of mass of the system relative to the sink, and the position vector of the star relative to its system's centre of mass. For binaries with long periods or very eccentric orbits, a star can be formed outside the sink if required by their orbital parameters.
Single stars inherit the velocity of the sink at the time of formation, plus a random fraction of the local sound speed~\citep{wall1}. We adopt this prescription for the systems' centres of mass. For stars in a binary system, their velocity is obtained from the addition of the sink's velocity, the random component from the local sound speed, and the velocity with respect to the centre of mass velocity. 

\section{Simulations}\label{sec:simulations}
Initial tests of \texttt{Torch}~\citep{wall1, wall2} have shown that the time evolution of the star formation rate and the spatial distribution of star formation are highly stochastic, and depend strongly on the initial conditions. We therefore adopt a single set of initial conditions for our full suite of simulations, to investigate solely the impact that the presence or absence of primordial binaries has on the final distributions of binaries.
We initialize all our simulations from the same spherical cloud of dense molecular gas with a mass of $10^4$ M$_{\odot}$, a virial parameter of 0.2, a radius of 7~pc and a Gaussian density profile with central density 8.73 x 10$^{-22}$ g/cm$^3$. 
The initial cloud has a central temperature of 20.64 K and sits in a medium of warm gas with a temperature of 6.11 x $10^{3}$ K and a density of 2.18 x 10$^{-22}$ g/cm$^3$. 
Each simulation uses the same initial turbulent Kolmogorov velocity spectrum but a different random seed for the stellar masses. There is no initial magnetic field. The gas follows an adiabatic equation of state with $\gamma = 5/3$. The simulations include atomic, molecular and dust cooling, as well as ionization, following~\citet{wall2}.

Galactic effects are ignored, as the clusters are evolved for $t \lesssim 3.2$ Myr. Tidal perturbations or disk crossing effects are unlikely to have an impact on the cluster's structure on such a timescale~\citep[e.g.][]{Kruijssen2011, Miholics2017}.
The size of the simulation box ($\sim 18$ pc) is large enough to ensure the choice of boundary conditions does not have a strong impact on the outcome of the simulation: observed star clusters in nearby galaxies with the same stellar mass as our simulations have half-mass radii one order of magnitude smaller than the box size~\citep[][and references therein]{Krumholz2019}. We use zero-gradient boundary conditions, which allow the gas and stars to leave the domain.
The choice of spatial resolution ($\sim 0.05$ pc) is appropriate to model the gas dynamics in the cluster, excluding star formation which is treated by a sub-grid model. The resolution is approximately one order of magnitude smaller than the average separation between stars in dense clusters~\citep[][and references therein]{Krumholz2019} and thus resolves well the behaviour of the gas between the stars.

We perform a total of 15 simulations, at two different maximum \texttt{FLASH} refinement levels. At our lowest refinement level, we perform five simulations with primordial binaries and five without primordial binaries; at our highest refinement level, we perform four simulations with primordial binaries and one simulation without primordial binaries. The least resolved regions in all our simulations are at refinement level 4 and have a gas spatial resolution of 0.136 pc. 
The spatial and mass resolutions of our simulations are presented in Table~\ref{tab:resolution}. 
In our analysis, we use the combined results of groups of simulations to ensure we have a large population of systems to analyze. We will use the variation between simulations to quantify the uncertainty in our results and the numerical effects of resolution.
We denote our suites of simulations with primordial binaries at refinement levels 5 and 6 as respectively \texttt{M4r5b} and \texttt{M4r6b}, and our suite of simulations without primordial binaries at refinement level 5 as \texttt{M4r5s}. We refer to our full suites of simulations at refinement levels 5 and 6 as respectively \texttt{M4r5} and \texttt{M4r6}; similarly, we refer to our full suite of simulations with primordial binaries as \texttt{M4b} and to our full suite of simulations without primordial binaries as \texttt{M4s}. We perform our analysis with 9866 stars in \texttt{M4r5b}, 9016 stars in \texttt{M4r5s}, 6384 stars in \texttt{M4r6b} and 1517 stars in \texttt{M4r6s}. 
We plot the results from \texttt{M4r5}, as this suite of simulations has the most stars.

We summarize the time of onset of star formation, the time at which the simulation is ended, the maximum stellar mass, the number of stars and the total stellar mass for each of our simulations in Table~\ref{tab:sf_times}. Since the only difference between the different simulations at the same resolution is in the stellar sampling, star formation starts at the same time and the first sink forms at the same location for all simulations at the same resolution. We present two examples of the time evolution of the star formation rate in Figure~\ref{fig:SFR}. In Figure~\ref{fig:frames},
we present the projected density
for nine simulations, a minimum of $\sim 1.5$ Myr after the onset of star formation. We note that the general structure of the gas and the sink locations are very similar in all simulations, as expected since all the simulations start from the same initial gas conditions. Nevertheless, the number of stars and their locations, as well as the total stellar mass, differ among the simulations. At similar times, there are spreads of 18\% in number of stars and 24\% in stellar mass. Our simulations end at 1.5 -- 2 Myr after the start of star formation, at the time when feedback from massive stars starts to have a significant impact on the gas properties. Therefore, our simulations probe the earliest stages of star cluster formation, when the dominant physical effects are gas collapse and star formation, combined with dynamical interactions between stars, binary systems, and their natal gas.

\begin{figure}
	\includegraphics[width=\columnwidth, clip=true, trim= 0.4cm 0cm 0.4cm 0cm]{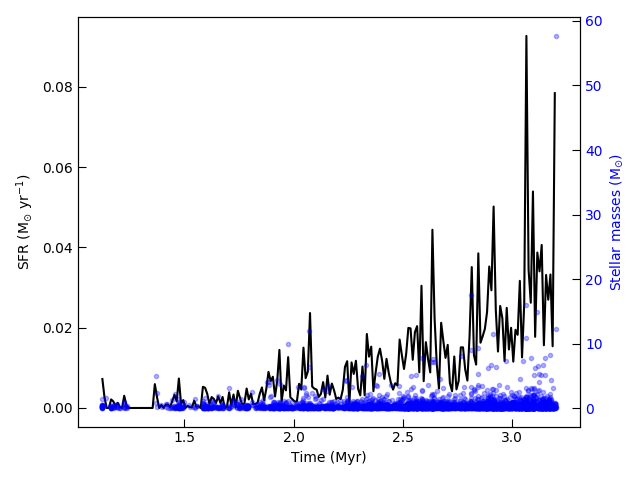}
	\includegraphics[width=\columnwidth, clip=true, trim= 0.4cm 0cm 0.4cm 0cm]{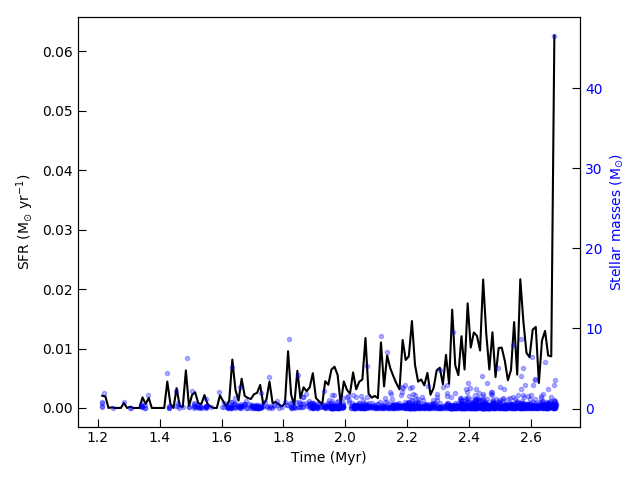}
    \caption{Star formation rates for \texttt{M4r5b-4} (top) and \texttt{M4r6s} (bottom). The solid black line shows the rate smoothed over 1 kyr (left axis) and the blue points shows the masses of the individual stars formed in the simulation (right axis). The total stellar masses are respectively 2.14 x 10$^{3}$ M$_{\odot}$ and 7.60 x 10$^{2}$ M$_{\odot}$ for \texttt{M4r5b-4} and \texttt{M4r6s}. Peaks in star formation rate coincide with the formation of massive stars, and there is an overall increase of the star formation rate as the simulation progresses.}
    \label{fig:SFR}
\end{figure}

\begin{figure*}
    \centering
    \includegraphics[clip=true, trim= 1.8cm 0 0.8cm 10cm, width=\linewidth]{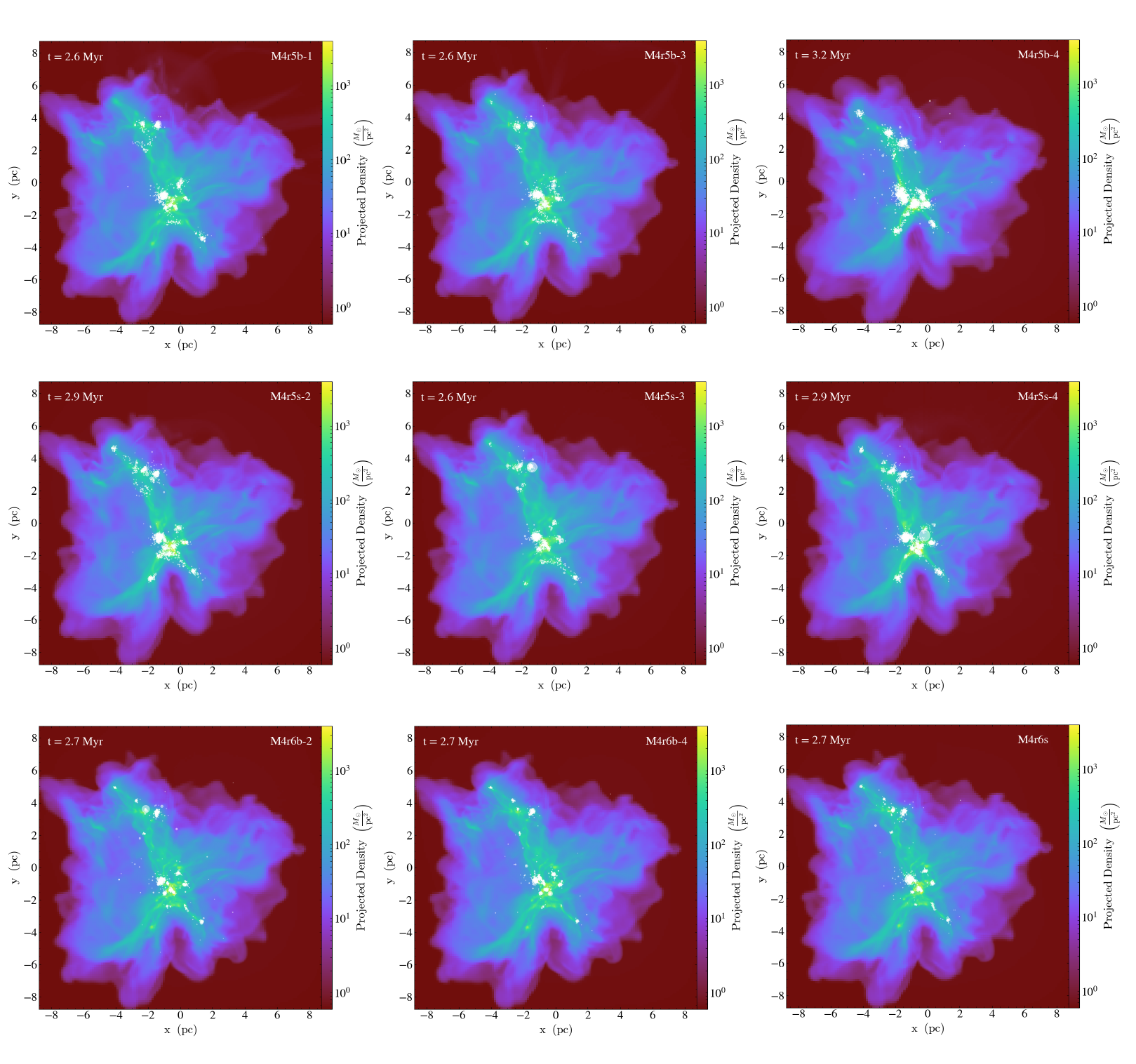}
    \caption{Final projected density distribution in the simulations (from top left to bottom right) \texttt{M4r5b-1}, \texttt{M4r5b-3}, \texttt{M4r5b-4}, \texttt{M4r5s-2}, \texttt{M4r5s-3}, \texttt{M4r5s-4}, \texttt{M4r6b-2}, \texttt{M4r6b-4} and \texttt{M4r6s}. The white circles are the full sample of stars in the simulations; the radius of the circle is proportional to the stellar mass. All simulations use the same initial conditions, which is reflected by the very similar gas configurations. We note however differences in the locations and masses of the stars, and expect to see these differences reflected in the gas configuration once feedback becomes more important. The total stellar mass and number of stars for each simulation can be found in Table~\ref{tab:sf_times}.}
    \label{fig:frames}
\end{figure*}

\begin{table}
	\centering
	\caption{Spatial and mass resolution, at maximum refinement level \textit{ref}. $\Delta x$ denotes the minimum zone size while $\Delta m$ and $\rho_c$ denote respectively the maximum mass and the maximum density in a grid cell to trigger sink formation, assuming a sound speed $c_s = 1.9$ x 10$^{4}$ cm s$^{-1}$~\citep[following][]{Federrath2010}}.
	\label{tab:resolution}
	\begin{tabular}{ccccc}
		\textit{ref} & $\Delta x$ (pc) & \textit{Sink diameter} (AU) & $\Delta m$ (M$_{\odot}$) & $\rho_{c}$ (g cm$^{-3}$) \\
		\hline
        5 & 6.83 x 10$^{-2}$ & 7.05 x 10$^{4}$ & 1.80 x 10$^{-2}$ & 3.82 x 10$^{-21}$ \\
        6 & 3.42 x 10$^{-2}$ & 3.53 x 10$^{4}$ & 9.00 x 10$^{-3}$ & 1.53 x 10$^{-20}$ \\
		\hline
	\end{tabular}
\end{table}

\begin{table}
	\centering
	\caption{Simulations. All simulations have a total gas mass $10^4$ M$_{\odot}$ and a minimum refinement of 4. The number after \texttt{r} denotes the maximum refinement level and the last letter indicates if the simulation includes primordial binaries (\texttt{b}) or single stars only (\texttt{s}). $t_{*}$ denotes the time of the onset of star formation and $t$ denotes the time at which the simulation has ended; $M_{m}$ denotes the mass of the most massive star, $N_*$ is the number of stars in the simulation, and $M_*$ is the total stellar mass.}
	\label{tab:sf_times}
	\begin{tabular}{lccccc} 
		\textit{Name} & $t_{*}$ (Myr) & $t$ (Myr) & $M_{m}$ (M$_{\odot}$) & $N_*$ & $M_*$ (M$_{\odot}$)\\
		\hline
        \texttt{M4r5b-1} & 1.12 & 2.61 & 17.61 & 1575 & 706 \\
        \texttt{M4r5b-2} & 1.12 & 2.43 & 67.13 & 976 & 490 \\
        \texttt{M4r5b-3} & 1.12 & 2.64 & 40.25 & 1661 & 774 \\
        \texttt{M4r5b-4} & 1.12 & 3.20 & 57.65 & 4704 & 2143 \\ 
        \texttt{M4r5b-5} & 1.12 & 2.36 & 32.83 & 950 & 417 \\ 
        \hline
        \texttt{M4r5s-1} & 1.12 & 2.43 & 59.25 & 877 & 473 \\
        \texttt{M4r5s-2} & 1.12 & 2.91 & 22.87 & 2487 & 1204 \\
        \texttt{M4r5s-3} & 1.12 & 2.65 & 68.49 & 1493 & 806 \\
        \texttt{M4r5s-4} & 1.12 & 2.94 & 78.84 & 2719 & 1400 \\
        \texttt{M4r5s-5} & 1.12 & 2.64 & 68.49 & 1440 & 774 \\
        \hline
        \texttt{M4r6b-1} & 1.21 & 2.65 & 10.92 & 1749 & 685 \\
        \texttt{M4r6b-2} & 1.21 & 2.68 & 38.85 & 1650 & 734 \\
        \texttt{M4r6b-3} & 1.21 & 2.60 & 16,42 & 1531 & 610 \\
        \texttt{M4r6b-4} & 1.21 & 2.66 & 20.23 & 1454 & 659 \\
        \texttt{M4r6s} & 1.21 & 2.68 & 46.57 & 1517 & 760 \\ 
        \hline
	\end{tabular}
\end{table}

\section{Binary properties}\label{sec:discussion}

\begin{figure}
	\includegraphics[width=\columnwidth, clip=true, trim= 0.5cm 1.5cm 1cm 0.5cm]{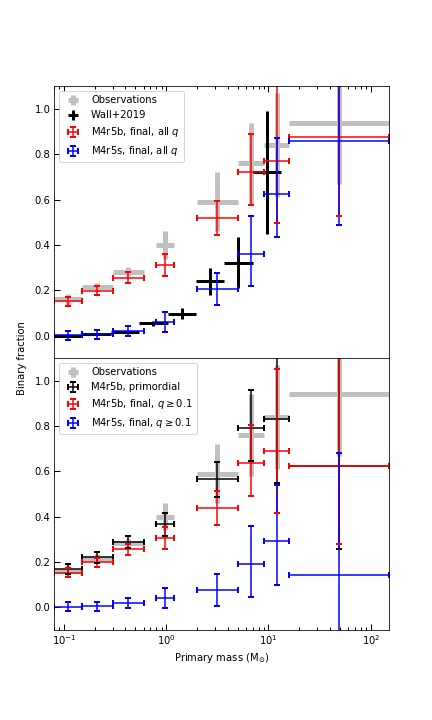}
    \caption{Binary fraction as a function of primary mass
    in \texttt{M4r5}, for the full binary population (top) and for observable systems (bottom).
    The primordial binaries formed in \texttt{M4r5b}, the binaries present at the end of \texttt{M4r5b} and those present at the end of \texttt{M4r5s} are denoted respectively by the black, red and blue thin crosses. The primordial and final binary fractions are exactly equal for the highest mass bin in the bottom panel. Observations from main sequence stars in the solar neighbourhood~\citep{Moe2017, winters}, with mass ratios $\geq 0.1$, are provided for comparison as the solid grey crosses. Binaries from the simulations of~\citet{wall1}, which do not include primordial binaries, are denoted by the thick black crosses. All the errors in $x$ correspond to the bin widths and the errors in $y$ in the observations are from the observational uncertainties. The $y$ errors on the simulation data from~\citeauthor{wall1} are $1/\sqrt{N}$~\citep[see][]{wall1}. The $y$ errors on our simulation data are from the Poisson statistical errors: $1/\sqrt{N}$ for $N > 100$ and the tabulated 1 $\sigma$ confidence interval for $N \leq 100$~\citep{Gehrels1986, measurements}.}
    \label{fig:fractions}
\end{figure}

As discussed in Section~\ref{subsec:mass_ratios}, observations of binary stars in the galactic field and in open clusters are only complete for mass ratios $q \geq 0.1$; consequently, our algorithm only generates primordial binaries with such mass ratios. Our work differs from previous studies of dynamical binary formation in clusters~\citep[e.g.][]{wall1} or of evolution of a population of primordial binaries in clusters~\citep[e.g.][]{Parker2014} by taking into account this observational limit, and comparing directly our population of binary systems with $q \geq 0.1$ to the observed field population. We emphasize that any system with mass ratio $q < 0.1$ would not have been included in the surveys on which our work is based. Where applicable, we therefore present two different sets of comparison: the comparison between our full simulation results and the field population (for consistency and ease of comparison with earlier literature), and the comparison between our simulation results with $q \geq 0.1$ (hereafter, observable simulation results) and the field population.

To consider stars to be members of a binary, we require the stars to be gravitationally bound and  perturbations by the other cluster stars must be comparatively small. Following~\cite{wall1}, we consider a system with primary mass $M_1$, companion mass $M_2$ and semi-major axis $a$ to be perturbed if there is a star with mass $M_p$ and within a distance $d$ of the system's centre of mass such that 
\begin{equation}
    \frac{4a^2}{M_1M_2} \Bigg|\frac{M_1 M_p}{(d-a)^2} - \frac{M_2 M_p}{(d+a)^2}\Bigg| > 3.
\end{equation}
To avoid considering stable triple systems as perturbed binaries, we add a condition that the system will not be considered perturbed if the third star is gravitationally bound to either the primary or the companion. Our conclusions are robust to the addition of this condition. To account for possible triple or higher order systems, we calculate the binary fraction as a function of primary mass as the fraction of stars in each mass range that are primaries (i.e. the most massive star in a stable system) but include each primary-companion pair in our analysis of binary properties.

\subsection{Binary fractions}
We compare the binary fraction from observations of main sequence field stars~\citep{Moe2017, winters}, the fraction of stars we form in primordial binaries, and the fraction of stars in unperturbed binary systems at the end of \texttt{M4b} and \texttt{M4s}. We plot the binary fraction as a function of primary mass for our full binary population (all $q$'s) and for observable systems ($q \geq 0.1$) in \texttt{M4r5} in Figure~\ref{fig:fractions}. 

We first consider our observable simulation results, which only include binary systems with $q \geq 0.1$. As expected, primordial binaries generated with our algorithm result in field-like binary fractions at all masses (see Figure~\ref{fig:algorithm_fraction}). The final distribution in \texttt{M4b} is consistent within uncertainties with observations and primordial fractions at all primary masses; we nevertheless note that the final fraction tends to be lower than either the observations or the primordial fraction. This trend is present at both resolutions; in \texttt{M4r6b}, the observable binary fraction between 0.30 and 0.60 M$_{\odot}$ is lower than what would be predicted by observations of main sequence field stars. This indicates that some primordial binaries are destroyed by dynamical interactions during cluster formation.
The final distribution in \texttt{M4s} is not consistent within uncertainties with observations, at any primary mass. It is however consistent within uncertainties with the primordial and final distributions in \texttt{M4b} for the two highest mass bins, where uncertainties are very large. 

For the full population of binary systems, we reproduce the results from~\cite{wall1} and find that, at high primary masses, pure dynamical formation results in binary fractions consistent with observations of main sequence field stars. We also find that with our field-like prescription for primordial binaries, our full binary population is consistent with observed binary fractions at all primary masses. We emphasize that these results should be used for comparison with previous literature, but do not reflect what we can observe due to the $q \geq 0.1$ detection limit in solar-neighbourhood surveys. 

Although pure dynamical formation leads to observable binary fractions consistent within uncertainties with observations at high primary masses, we argue that this is due only to the large uncertainties arising from the very small number of stars in the highest mass bins. 
In contrast with~\cite{wall1}, we find that we need primordial binaries at all primary masses in order to be consistent with observations of main sequence field stars, due to the additional constraint that our systems must have large enough mass ratios to be seen in observations. 

\begin{figure}
	\includegraphics[width=\columnwidth, clip=true, trim= 0.5cm 1.5cm 1cm 0.5cm]{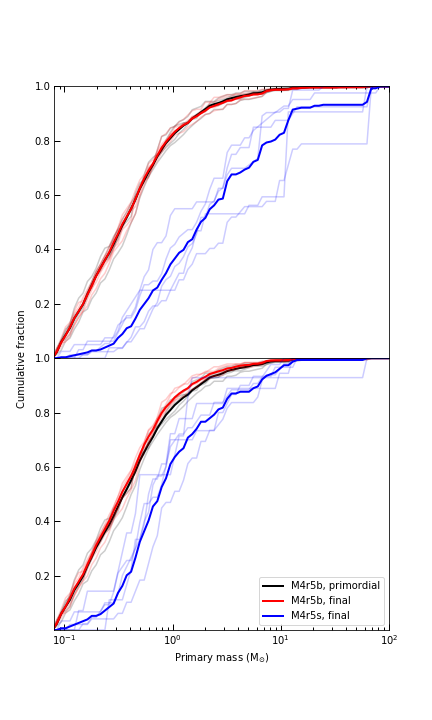}
    \caption{Cumulative distribution of primary masses in \texttt{M4r5}, for the full binary population (top) and observable systems (bottom). The solid black line denotes the primordial primary mass distribution in \texttt{M4r5b}, the solid red line denotes the final distribution in \texttt{M4r5b} and the solid blue line denotes the final distribution in \texttt{M4r5s}. The fainter lines denote the corresponding primary mass distribution in individual simulations. Pure dynamical formation results in systems with higher primary masses, while the dynamical evolution of the cluster with primordial binaries favours lower-mass primaries.}
    \label{fig:sims_primary_mass}
\end{figure}

\begin{figure}
	\includegraphics[width=\columnwidth, clip=true, trim= 0.5cm 1.5cm 1cm 0.5cm]{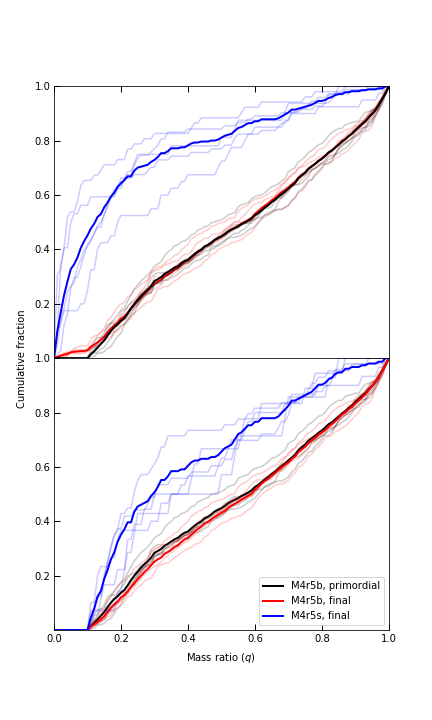}
    \caption{Cumulative distribution of mass ratios in \texttt{M4r5}, for the full binary population (top) and observable systems (bottom). The solid black line denotes the primordial mass ratio distribution in \texttt{M4r5b}, the solid red line denotes the final distribution in \texttt{M4r5b} and the solid  blue line denotes the final distribution in \texttt{M4r5s}. The fainter lines denote the corresponding mass ratio distribution in individual simulations. Dynamical formation in both \texttt{M4r5b} and \texttt{M4r5s} favours systems with smaller mass ratios: up to 50\% of the systems formed in \texttt{M4r5s} have mass ratios below the detection limit.
    }
    \label{fig:sims_mass_ratio}
\end{figure}

\begin{figure}
	\includegraphics[width=\columnwidth, clip=true, trim= 0.5cm 1.5cm 1cm 0.5cm]{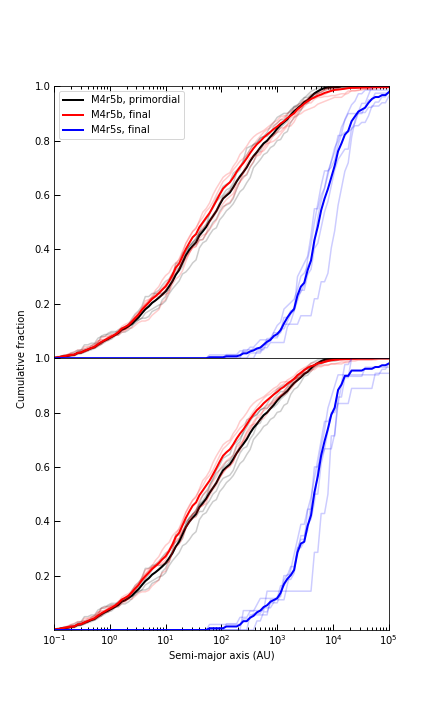}
    \caption{Cumulative distribution of semi-major axes in \texttt{M4r5}, for the full binary population (top) and observable systems (bottom). The solid black line denotes the primordial semi-major axis distribution in \texttt{M4r5b}, the solid red line denotes the final distribution in \texttt{M4r5b} and the solid blue line denotes the final distribution in \texttt{M4r5s}. The fainter lines denote the corresponding semi-major axis distribution in individual simulations. Pure dynamical formation results in systems with much larger semi-major axes; conversely, dynamical evolution during cluster formation results in smaller semi-major axes.}
    \label{fig:sims_semi_major_axis}
\end{figure}

\begin{figure}
	\includegraphics[width=\columnwidth, clip=true, trim= 0.5cm 1.5cm 1cm 0.5cm]{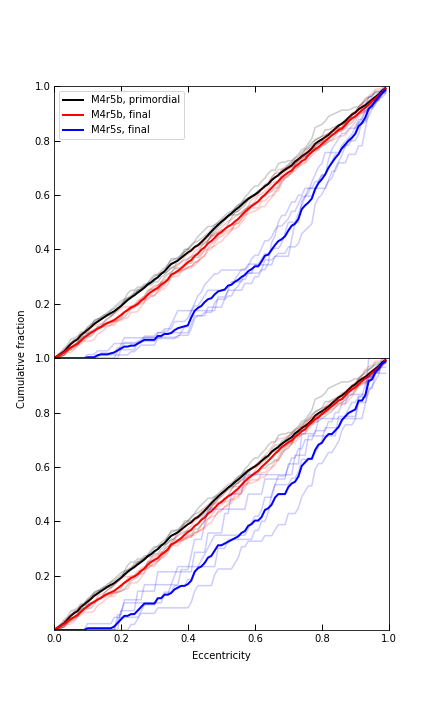}
    \caption{Cumulative distribution of eccentricities in \texttt{M4r5}, for the full binary population (top) and observable systems (bottom). The solid black line denotes the primordial eccentricity distribution in \texttt{M4r5b}, the solid red line denotes the final distribution in \texttt{M4r5b} and the solid blue line denotes the final distribution in \texttt{M4r5s}. The fainter lines denote the corresponding eccentricity distribution in individual simulations. Pure dynamical formation results in systems with much larger eccentricities.}
    \label{fig:sims_ecccentricity}
\end{figure}

\begin{figure}
	\includegraphics[width=\columnwidth, clip=true, trim= 0.5cm 1.5cm 1cm 0.5cm]{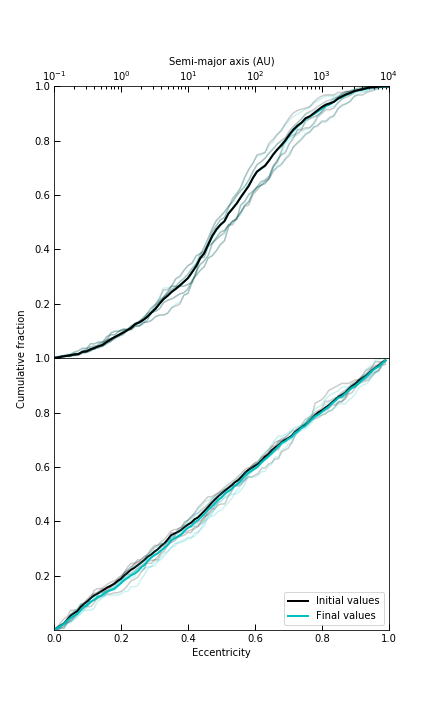}
    \caption{Cumulative distributions of semi-major axes (top) and eccentricities (bottom) for the primordial systems in \texttt{M4r5b} surviving to the end of our simulations. The solid grey lines represent the initial properties of the surviving systems, while the solid cyan lines represent their final properties. The fainter lines denote the corresponding distributions in individual simulations. Out of the 1274 binary systems we detect in \texttt{M4r5b}, 1077 systems are surviving systems. The distribution of the semi-major axes of the surviving systems at the end of the simulations is consistent with their distribution at the time of star formation (96.2\% confidence). The final values of eccentricity are systematically larger (83.1\% confidence). By definition, the primary masses and mass ratios of these systems are unchanged. 
    }
    \label{fig:sims_change_in_properties}
\end{figure}

\begin{figure*}
	\includegraphics[width=\linewidth, clip=True, trim = 4.5cm 0 5cm 0]{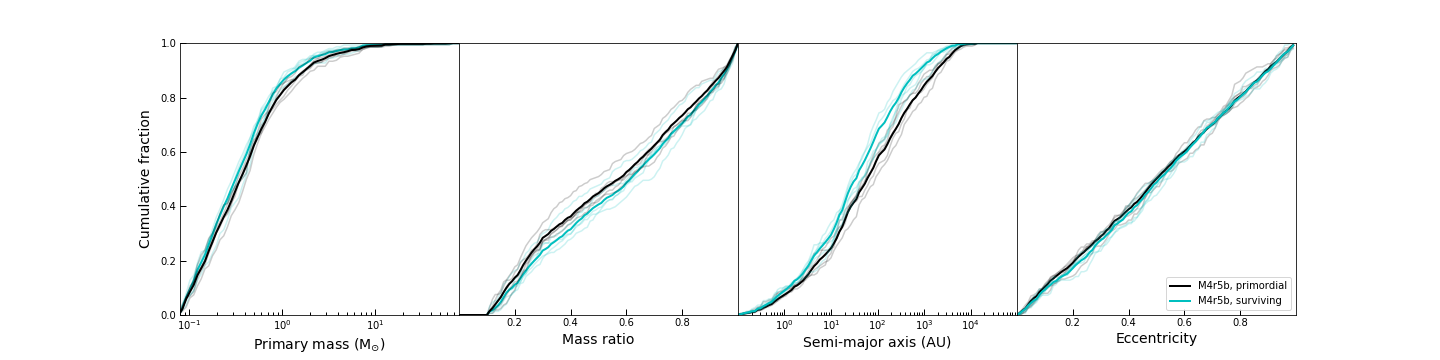}
    \caption{Cumulative distributions of primary masses, mass ratios, semi-major axes and eccentricities for the primordial systems in \texttt{M4r5b}. The black solid lines represent all the primordial systems formed in our simulations and the solid cyan lines represent the primordial systems that survive until the end of simulations. The fainter lines denote the corresponding distributions in individual simulations. Out of 1789 primordial systems, 260 were fully destroyed and 66 changed companions; 1463 systems survived to the end of our simulations. The distributions of primary masses, mass ratios and semi-major axes are different for the full primordial population and the subset of surviving systems; it is ambiguous whether the eccentricity distribution changed.}
    \label{fig:sims_surviving_vs_primordial}
\end{figure*}

\begin{figure*}
	\includegraphics[width=\linewidth, clip=True, trim = 4.5cm 0 5cm 0]{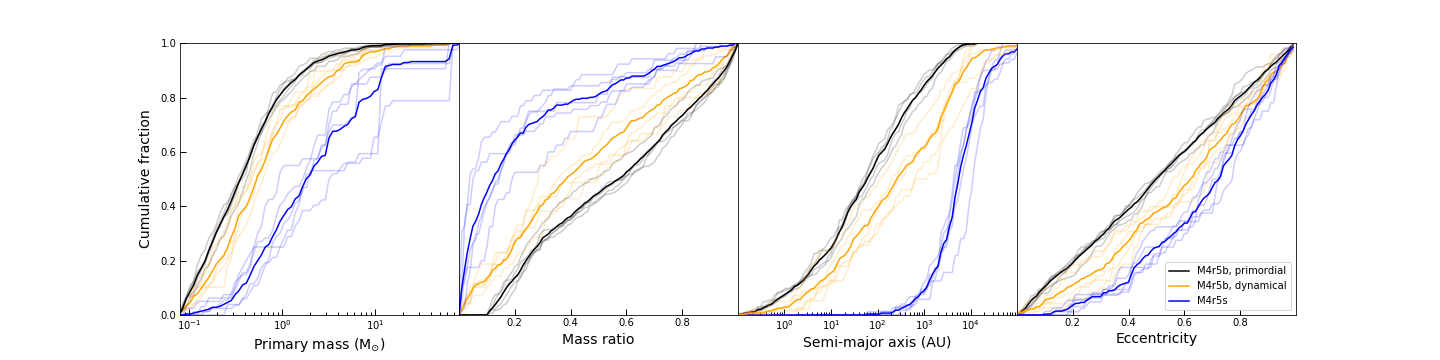}
    \caption{Cumulative distributions of primary masses, mass ratios, semi-major axes and eccentricities for binary systems formed dynamically in \texttt{M4r5}, and primordial binaries in \texttt{M4r5b}. The solid blue lines represent the 281 systems detected in \texttt{M4r5s}, the solid orange lines represent the 66 systems where the primary changed companions and the 341 systems formed dynamically with new primaries in \texttt{M4r5b}, and the solid black lines represent the primordial binaries in \texttt{M4r5b}. The fainter lines denote the corresponding distributions in individual simulations. The distributions are different for all four parameters; we note that the cumulative distributions for dynamical binaries in \texttt{M4r5b} always lie between the cumulative distributions for primordial binaries and \texttt{M4r5s}.}
    \label{fig:sims_dynamical_properties}
\end{figure*}

\subsection{Final binary properties}
We compare the final distributions of primary masses, mass ratios, semi-major axes and eccentricities in our simulations, and test the null hypothesis that they are drawn from the same distribution with the Mann-Whitney-Wilcoxon $U$-test~\citep{Wilcoxon1945, Mann1947}, which is similar to the Kolmogorov-Smirnov test but more suitable for larger numbers of data points. 
We consider the primordial distributions, and the full and observable final distributions in \texttt{M4b} and \texttt{M4s}. Where relevant, we quote the lowest confidence level we have between \texttt{M4r5} and \texttt{M4r6}. The qualitative conclusions are always the same at both resolutions for observable systems. When comparing the primordial and final distributions, we find that our conclusions hold for each individual simulation.

We present the plot of the cumulative primary mass distributions for \texttt{M4r5} in Figure~\ref{fig:sims_primary_mass}. 
We do not reject the null hypothesis that the primordial and full final distributions of primary masses in \texttt{M4b} are drawn from the same underlying distribution. If we consider only observable systems, however, we find that the systems detected at the end of \texttt{M4b} have lower primary masses than the systems formed primordially in these simulations (93.5\% confidence).
Furthermore, primary masses at the end of \texttt{M4s} are higher than in \texttt{M4b}, both at the beginning and end of the simulations ($> 99.9\%$ confidence).

We plot the cumulative mass ratio distributions for \texttt{M4r5} in Figure~\ref{fig:sims_mass_ratio}. The mass ratios for the full binary distributions at the end of \texttt{M4b} are consistent with having been drawn from the same distribution as the primordial mass ratios. This result is in agreement with previous studies of binaries in clusters~\citep[e.g.][]{Parker2013}. Mass ratios of observable systems, however, are larger than primordial mass ratios (91.6\% confidence). 
This alteration of the mass ratio distribution is in agreement with the results from simulations of young~\citep[e.g.][]{Parker2012} and globular~\citep[e.g.][]{Sollima2008} clusters. Mass ratios in \texttt{M4s} are smaller than those in \texttt{M4b}, either at the time of star formation or at the end of the simulation ($>99.9\%$ confidence). 

The cumulative semi-major axis distributions for \texttt{M4r5} are shown in Figure~\ref{fig:sims_semi_major_axis}.
We find that the semi-major axes of systems detected at the end of \texttt{M4b} are smaller than those of the primordial systems. We are confident at respectively 96.9\% and $> 99.9\%$ that it is the case for our full sample of systems, and our sub-sample of observable systems. Conversely, the systems in \texttt{M4s} have larger semi-major axes than those formed primordially or those detected at the end of \texttt{M4b} ($> 99.9\%$ confidence). This suggests that systems with large semi-major axes are preferentially formed dynamically.

We also plot the cumulative distribution of eccentricities in \texttt{M4r5} in Figure~\ref{fig:sims_ecccentricity}. 
The systems detected at the end of \texttt{M4b} are more eccentric than the primordial systems formed in the simulation for either our full sample (99.6\% confidence) or just the observable systems (98.3\%). 
This result is consistent with what we would expect of long-term dynamical evolution of binary systems in clusters, where repeated dynamical encounters increase eccentricities~\citep[e.g.][]{Hills1975, Heggie1996, Ivanova2006}.
Similarly, we are also confident at $>99.9\%$ that systems in \texttt{M4s} have larger eccentricities than either those formed primordially or those detected at the end of \texttt{M4b}. 
We argue that dynamical interactions form eccentric systems preferentially, causing the larger eccentricities in \texttt{M4b} and especially \texttt{M4s}.

All the changes we detect in the distributions are small but statistically significant. They suggest that very early during cluster formation, while there is still a significant amount of gas and active star formation, dynamical interactions between the stars already modify binary systems in a non-random way. This highlights the need for the concurrent inclusion of gas and binaries in star cluster formation and early evolution simulations.

\subsection{Modification of primordial binaries}
We investigate the fate of the systems formed primordially in our simulations. We present the cumulative distributions of semi-major axes and eccentricities for surviving systems (i.e. systems that have the same companion at the time of formation and at the end of the simulation) in \texttt{M4r5} in Figure~\ref{fig:sims_change_in_properties}. For \texttt{M4r5b} and \texttt{M4r6b}, we are confident at respectively 96.2\% and 85.2\% that the primordial and final semi-major axes are drawn from the same distribution. We are also confident at 83.1\% and 89.7\% that the surviving systems are more eccentric at the end of the simulation than when they form. This change in eccentricity of the surviving systems is consistent with our earlier result that systems at the end of \texttt{M4b} tend to have larger eccentricities than the primordial systems. Despite this result, the changes in the eccentricity distribution are very small, and are unlikely to be dynamically significant. We would expect long term evolution to cause an increase in eccentricity of hard binaries through dynamical interactions~\citep{Hills1975, Heggie1996, Ivanova2006}; we may see here the beginnings of this phenomenon.

We also compare the properties of the surviving subset of primordial systems to those of the full primordial population. 
The relevant primary masses, mass ratios, semi-major axes and eccentricities are plotted in Figure~\ref{fig:sims_surviving_vs_primordial}. We quote our lowest confidence level between \texttt{M4r5b} and \texttt{M4r6b}, and verify that our conclusions hold for the surviving primordial binaries in any individual simulation. 
We are confident at 99.6\% that the primordial systems surviving to the end of the simulation have smaller primary masses than the full primordial population. We are also confident at 98.5\% that surviving primordial systems have larger mass ratios than the full population of primordial systems. This suggests that systems with high primary masses and small mass ratios are the most likely to either be dynamically destroyed or change companion due to three-body interactions.

We are confident at $>99.9\%$ that surviving primordial systems have smaller semi-major axes than the full primordial distribution. As the semi-major axes of surviving systems are not modified, we attribute this to the preferred dynamical destruction or modification of systems with large semi-major axes, as expected from the Heggie-Hills Law~\citep{Heggie1975, Hills1975}.
Finally, we are confident at 67.9\% that surviving primordial systems have larger eccentricities than the full sample of primordial systems. On its own, this result does not indicate that systems with smaller eccentricities are preferentially destroyed, as it appears that the eccentricities of surviving systems increase during the simulation.
Our results indicate that primordial binaries are destroyed within our simulations, and that systems may be more likely to be destroyed if they have certain properties. They also suggest that the orbital properties of primordial systems may already be modified in the earliest stages of cluster formation.

\subsection{Dynamical binary formation}

\begin{figure*}
	\includegraphics[width=\linewidth, clip=True, trim = 4.5cm 0 5cm 0]{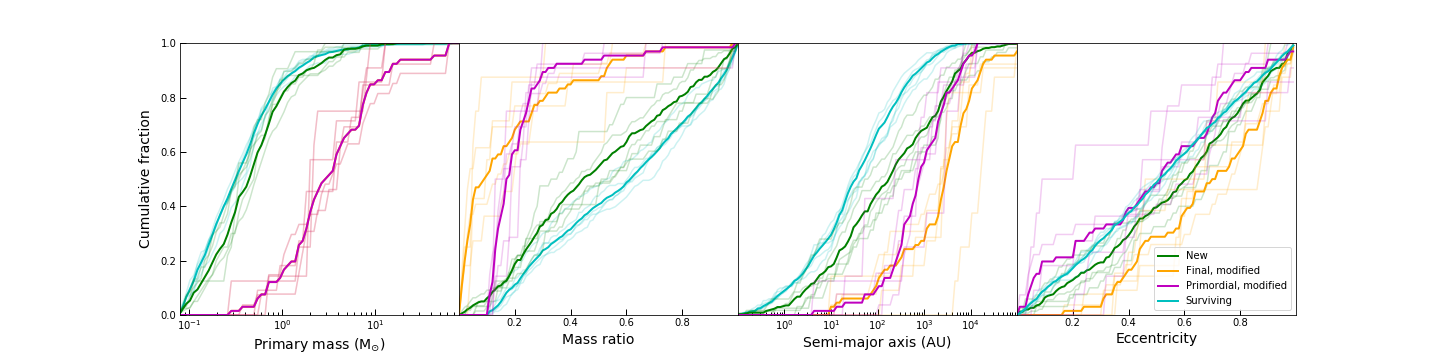}
    \caption{Cumulative distributions of primary masses, mass ratios, semi-major axes and eccentricities for surviving, modified and new binary systems in \texttt{M4r5b}. The solid green lines represent the 341 systems formed dynamically with new primaries in \texttt{M4r5b}, the solid magenta and orange lines represent the 66 systems where the primary changed companions respectively when they are formed and at the end of \texttt{M4r5b}, and the solid cyan lines represent the surviving systems in \texttt{M4r5b}. The fainter lines denote the corresponding distributions in individual simulations. The distributions are all different except for the primary mass distribution, which is by definition the same for the modified systems when they are formed and at the end of the simulation.}
    \label{fig:sims_dynamical_change}
\end{figure*}

We also investigate whether the properties of the binaries formed dynamically in \texttt{M4b} and \texttt{M4s} are the same. We present plots of the primary masses, mass ratios, semi-major axes and eccentricities for binaries formed dynamically in \texttt{M4r5} in Figure~\ref{fig:sims_dynamical_properties}. In \texttt{M4b}, we consider that a binary is formed dynamically if the primary changed companion or the primary was not previously in a binary system. 

We find, with confidence $> 99.9\%$, that the properties of binaries formed dynamically in simulations including primordial binaries are different from either the properties of primordial binaries or the properties of binaries formed dynamically in a simulation without primordial binaries. Furthermore, we find that the cumulative distributions of properties of binaries formed dynamically in \texttt{M4b} always lie between the cumulative distributions for primordial binaries and for \texttt{M4s}. Binaries formed dynamically in the presence of primordial binaries tend to have smaller primary masses than those arising from pure dynamical formation, but larger primary masses than primordial binaries. Conversely, they tend to have mass ratios smaller than the systems formed by pure dynamical interactions, but larger than the primordial systems.
In \texttt{M4b}, dynamical binaries form with semi-major axes and eccentricities larger than the primordial systems, but smaller than they do in \texttt{M4s}.

These early results are consistent with our expectations of dynamical formation of binaries. In the case where binaries are formed primarily though single-single interactions, the resultant systems are more likely to have large primary masses, wide separations, small mass ratios, and large eccentricities as seen in Figure \ref{fig:sims_dynamical_properties}. When binaries can be formed through single-binary or higher order interactions, more complicated outcomes occur. High-mass primaries are still favoured, but systems with lower primary masses and high total mass (i.e. high mass ratio) are also likely to be involved in a dynamical encounter. During higher-order encounters, the rule of thumb is that the lowest-mass object involved in the encounter is ejected and replaced with a higher-mass object \citep{1993ApJ...415..631S}. Therefore, we might expect that dynamical encounters would tend to shift the mass ratio distribution towards larger values: the ejection of the lowest-mass object would result in an increase of a system's mass ratio following each higher-order encounter. What we see in our simulations is that the systems which lose their original companion and later gain another one are typically high mass systems with comparatively very small companions at large semi-major axes, as shown in Figure~\ref{fig:sims_dynamical_change}. These systems tend to replace their original companion by a lower-mass one, which goes against our expectations for higher-order encounters. However, we see evidence that some of these weakly-bound binaries are broken up very soon after formation: some primaries lose their original companion within the first 0.1 Myr after they are formed. The massive primary then essentially acts like a single star, capturing low-mass single stars on wide eccentric orbits, or exchanging into a lower-mass binary system. 

We also create new binaries from primaries which were originally single stars. Those primaries also have slightly higher masses than the underlying population, and become binaries by capturing a lower-mass single star or exchanging into a binary system.  Tighter binary systems with smaller semi-major axes can be created through these exchange encounters than in the single-single case, and the extreme eccentricities that come from the near-parabolic encounters are not necessary when one of the original systems is already a binary.

\section{Results}\label{sec:conclusion}
We implement primordial binaries in the coupled MHD and direct N-body code \texttt{Torch}~\citep{wall1}, which couples \texttt{FLASH}~\citep{Fryxell2000} with the N-body code \texttt{ph4}~\citep{McMillan2012} and the stellar evolution code \texttt{SeBa}~\citep{seba} via the \texttt{AMUSE} framework~\citep{amuse}. 
We develop an algorithm to generate a population of binaries with mass-dependent binary fractions, periods, mass ratios and eccentricities. We also modify the star formation routine in \texttt{Torch} to force the concurrent formation of the stars in a binary system.
As an ansatz, we use the field distribution as our initial population of binaries. 
We perform 15 simulations; after the initial onset of star formation in each simulation, we see continuous and increasing star formation. Nine of our simulations include a population of primordial binaries, introduced following an extensive set of prescriptions. We follow the dynamical evolution of the binary population, and characterize it at the end of the simulations, 1.2--2 Myr after the onset of star formation. These first results suggest that concurrently modelling gas, stellar dynamics and binary systems during the earliest stages of star cluster formation is important, as binary systems are already being modified.

We investigate the impact of dynamical interactions during cluster formation on the primordial population of binaries. Our results indicate that dynamical interactions cause small but statistically significant changes in the distributions of binaries' primary masses, mass ratios, semi-major axes and eccentricities for systems above the $q \geq 0.1$ detection limit. We note that if we consider the full binary population (i.e. if we also consider systems with mass ratios $q < 0.1$), the differences in the distributions of primary masses and mass ratios are not obvious. We also find that primordial binaries are needed at all primary masses to reproduce the observed binary fraction above the $q \geq 0.1$ detection limit.
We argue that the distinction between the full binary population and the subset of observable systems is important, as observations are incomplete for $q < 0.1$ and considering only the systems with $q \geq 0.1$ significantly affects our conclusions. We find that all our conclusions are robust to a change in spatial resolution by a factor of 2.

We observe both dynamical formation and destruction of binary systems in \texttt{M4b}, which includes an initial population of binaries. In these simulations, we see that systems formed dynamically do not have the same properties as primordial systems, and more importantly, that systems formed dynamically in \texttt{M4b} do not have the same properties as those formed in \texttt{M4s}, which includes only single stars initially. The cumulative distributions of primary masses, mass ratios, semi-major axes and eccentricities formed dynamically in \texttt{M4b} lie between the primordial distribution and the distribution resulting from pure dynamical formation in \texttt{M4s}. The presence of an initial population of binary stars has a significant impact on the subsequent binary properties in the star cluster.
We find that systems with higher primary masses, lower mass ratios, larger semi-major axes and larger eccentricities are preferentially formed dynamically. We also find that systems with higher primary masses, smaller mass ratios and larger semi-major axes are preferentially destroyed or modified by dynamical interactions. Globally, dynamical evolution of a field-like primordial population favours systems with smaller primary masses, larger mass ratios, smaller semi-major axes and larger eccentricities. Most importantly, our results demonstrate that even in the earliest stages of cluster formation, when there is still a significant amount of gas and active star formation, dynamical interactions modify the binary population.

\section{Discussion}\label{sec:discussion2}
These simulations indicate that dynamical interactions in embedded clusters modify the properties of the primordial distribution of binaries by forming and destroying systems, but do not modify the mass-dependent binary fraction.
We emphasize that our simulations model the earliest stages of star cluster formation, and thus that we are probing those dynamical interactions that act on the binary systems on short timescales. Our analysis is conducted 1.2 -- 2 Myr after the onset of star formation, while there is still active star formation and there has been very little feedback from the stars. Furthermore, protostellar outflows, which we do not include, play a role in regulating star formation efficiency in low-mass star forming regions~\citep{Matzner2000}. 
With protostellar outflows, fewer stars would be formed during the earliest stages of cluster formation, and thus dynamical interactions between these stars would likely have a smaller impact on the properties of the binary distribution. Magnetic fields, which are absent from our simulations, also participate in the regulation of star formation~\citep{Price2008}.
Gas dynamical friction, which acts on scales smaller than our gas spatial resolution, may be a channel for the formation of short-period binaries with circular orbits~\citep{Gorti1996, Stahler2010}. Its absence may play a part in driving the shift towards larger semi-major axes and eccentricities. 
Our simulations were also conducted with a single choice for the initial gas properties (total mass, initial size of the cloud, etc). Additional simulations are needed to determine whether the global gas properties play a significant role in modifying the population of binary stars.

The next steps are to investigate the impact of an initial magnetic field on the evolution of an initial population of binaries, as well as the impact of stellar winds.
Massive stars will have a significant impact on the forming cluster: they interact gravitationally with other stars and deplete the supply of cold molecular gas available for star formation by increasing its temperature and ejecting it from the cluster.

In future work, we will alter our assumed primordial binary distribution to empirically determine what distribution leads to the field binary distribution observed after dynamical interactions in the embedded cluster.
An important feature of the dynamical evolution appears to be the destruction of systems with massive primaries, or the replacement of the observable companion by a companion with $q < 0.1$ in such systems. Our altered distribution should therefore favour the retention of the original companion in systems with massive primaries, which could be done by assuming smaller semi-major axes. This would be expected for primordial binaries forming from the fragmentation of a single core.
In addition, primordial binaries with mass ratios $q < 0.1$ likely do form primordially and may have a dynamically interesting effect on the binary populations. 
Similarly, our primordial binary population is based on the full distributions of parameters for observed primary-companion pairs in the field: the distributions include mass ratios and semi-major axes from the outer components of triples and higher order systems. Such systems are ubiquitous at high masses but the outer components are likely to have small mass ratios and large semi-major axes. An avenue to explore for our altered distribution would be to use distributions derived exclusively from only binaries and the inner components of hierarchical systems.
It is likely some of the systems detected in our simulations are dynamically formed stable triples or higher order systems, which we will also address in future work.

\section*{Acknowledgements}
We warmly thank Ralf Klessen for useful discussions. We also thank the referee, Douglas Heggie, for comments that improved the manuscript. 
We gratefully acknowledge the hospitality of the Centre for Computational Astrophysics, where this work was started during the first Torch users meeting in 2019. M-MML, SLWM, and AT are partly supported by US NSF grant AST18-15461. CCC and AS are supported by the Natural Sciences and Engineering Research Council of Canada. The simulations in this work were conducted on Cartesius; we acknowledge the Dutch National Supercomputing Center SURFSara grant 15520.

\section*{Data availability}
The data underlying this article will be shared on reasonable request to the corresponding author.

%%%%%%%%%%%%%%%%%%%%%%%%%%%%%%%%%%%%%%%%%%%%%%%%%%



\bibliographystyle{mnras}
\bibliography{references} % bibtex file is called references.bib

\begin{thebibliography}{}
\makeatletter
\relax
\def\mn@urlcharsother{\let\do\@makeother \do\$\do\&\do\#\do\^\do\_\do\%\do\~}
\def\mn@doi{\begingroup\mn@urlcharsother \@ifnextchar [ {\mn@doi@}
  {\mn@doi@[]}}
\def\mn@doi@[#1]#2{\def\@tempa{#1}\ifx\@tempa\@empty \href
  {http://dx.doi.org/#2} {doi:#2}\else \href {http://dx.doi.org/#2} {#1}\fi
  \endgroup}
\def\mn@eprint#1#2{\mn@eprint@#1:#2::\@nil}
\def\mn@eprint@arXiv#1{\href {http://arxiv.org/abs/#1} {{\tt arXiv:#1}}}
\def\mn@eprint@dblp#1{\href {http://dblp.uni-trier.de/rec/bibtex/#1.xml}
  {dblp:#1}}
\def\mn@eprint@#1:#2:#3:#4\@nil{\def\@tempa {#1}\def\@tempb {#2}\def\@tempc
  {#3}\ifx \@tempc \@empty \let \@tempc \@tempb \let \@tempb \@tempa \fi \ifx
  \@tempb \@empty \def\@tempb {arXiv}\fi \@ifundefined
  {mn@eprint@\@tempb}{\@tempb:\@tempc}{\expandafter \expandafter \csname
  mn@eprint@\@tempb\endcsname \expandafter{\@tempc}}}

\bibitem[\protect\citeauthoryear{{Abt} \& {Levy}}{{Abt} \&
  {Levy}}{1976}]{Abt1976}
{Abt} H.~A.,  {Levy} S.~G.,  1976, \mn@doi [\apjs] {10.1086/190363}, \href
  {https://ui.adsabs.harvard.edu/abs/1976ApJS...30..273A} {30, 273}

\bibitem[\protect\citeauthoryear{{Baczynski}, {Glover}  \&
  {Klessen}}{{Baczynski} et~al.}{2015}]{fervent}
{Baczynski} C.,  {Glover} S.~C.~O.,   {Klessen} R.~S.,  2015, \mn@doi [\mnras]
  {10.1093/mnras/stv1906}, \href
  {https://ui.adsabs.harvard.edu/abs/2015MNRAS.454..380B} {454, 380}

\bibitem[\protect\citeauthoryear{{Chabrier}}{{Chabrier}}{2003}]{Chabrier2003}
{Chabrier} G.,  2003, \mn@doi [\pasp] {10.1086/376392}, \href
  {https://ui.adsabs.harvard.edu/abs/2003PASP..115..763C} {115, 763}

\bibitem[\protect\citeauthoryear{{Chini}, {Hoffmeister}, {Nasseri}, {Stahl}  \&
  {Zinnecker}}{{Chini} et~al.}{2012}]{Chini2012}
{Chini} R.,  {Hoffmeister} V.~H.,  {Nasseri} A.,  {Stahl} O.,   {Zinnecker} H.,
   2012, \mn@doi [\mnras] {10.1111/j.1365-2966.2012.21317.x}, \href
  {https://ui.adsabs.harvard.edu/abs/2012MNRAS.424.1925C} {424, 1925}

\bibitem[\protect\citeauthoryear{{Colella} \& {Woodward}}{{Colella} \&
  {Woodward}}{1984}]{Colella1984}
{Colella} P.,  {Woodward} P.~R.,  1984, \mn@doi [J. Comput. Phys.]
  {10.1016/0021-9991(84)90143-8}, \href
  {https://ui.adsabs.harvard.edu/abs/1984JCoPh..54..174C} {54, 174}

\bibitem[\protect\citeauthoryear{{Deacon} \& {Kraus}}{{Deacon} \&
  {Kraus}}{2020}]{Deacon2020}
{Deacon} N.~R.,  {Kraus} A.~L.,  2020, \mn@doi [\mnras]
  {10.1093/mnras/staa1877}, \href
  {https://ui.adsabs.harvard.edu/abs/2020MNRAS.496.5176D} {496, 5176}

\bibitem[\protect\citeauthoryear{{Delfosse} et~al.,}{{Delfosse}
  et~al.}{2004}]{Delfosse2004}
{Delfosse} X.,  et~al., 2004, in {Hilditch} R.~W.,  {Hensberge} H.,
  {Pavlovski} K.,  eds,  ASP Conference Series Vol. 318, Spectroscopically and
  Spatially Resolving the Components of the Close Binary Stars. Astronomical
  Society of the Pacific, San Francisco, pp 166--174

\bibitem[\protect\citeauthoryear{{Duch{\^e}ne} \& {Kraus}}{{Duch{\^e}ne} \&
  {Kraus}}{2013}]{Duchene2013}
{Duch{\^e}ne} G.,  {Kraus} A.,  2013, \mn@doi [\araa]
  {10.1146/annurev-astro-081710-102602}, \href
  {https://ui.adsabs.harvard.edu/abs/2013ARA&A..51..269D} {51, 269}

\bibitem[\protect\citeauthoryear{{Duch{\^e}ne}, {Bouvier}, {Simon}, {Close}  \&
  {Eisl{\"o}ffel}}{{Duch{\^e}ne} et~al.}{1999}]{Duchene1999}
{Duch{\^e}ne} G.,  {Bouvier} J.,  {Simon} T.,  {Close} L.,   {Eisl{\"o}ffel}
  J.,  1999, in Bonaccini D.,  ed.,  ESO Conference and Workshop Proceedings
  Vol. 56, Astronomy with adaptive optics : present results and future
  programs. European Southern Observatory, Garching, p.~185

\bibitem[\protect\citeauthoryear{{Duch{\^e}ne}, {Lacour}, {Moraux}, {Goodwin}
  \& {Bouvier}}{{Duch{\^e}ne} et~al.}{2018}]{Duchene2018}
{Duch{\^e}ne} G.,  {Lacour} S.,  {Moraux} E.,  {Goodwin} S.,   {Bouvier} J.,
  2018, \mn@doi [\mnras] {10.1093/mnras/sty1180}, \href
  {https://ui.adsabs.harvard.edu/abs/2018MNRAS.478.1825D} {478, 1825}

\bibitem[\protect\citeauthoryear{{Duquennoy} \& {Mayor}}{{Duquennoy} \&
  {Mayor}}{1991}]{Duquennoy1991}
{Duquennoy} A.,  {Mayor} M.,  1991, \aap, \href
  {https://ui.adsabs.harvard.edu/abs/1991A&A...248..485D} {500, 337}

\bibitem[\protect\citeauthoryear{{Federrath}, {Banerjee}, {Clark}  \&
  {Klessen}}{{Federrath} et~al.}{2010}]{Federrath2010}
{Federrath} C.,  {Banerjee} R.,  {Clark} P.~C.,   {Klessen} R.~S.,  2010,
  \mn@doi [\apj] {10.1088/0004-637X/713/1/269}, \href
  {https://ui.adsabs.harvard.edu/abs/2010ApJ...713..269F} {713, 269}

\bibitem[\protect\citeauthoryear{{Fischer} \& {Marcy}}{{Fischer} \&
  {Marcy}}{1992}]{Fischer1992}
{Fischer} D.~A.,  {Marcy} G.~W.,  1992, \mn@doi [\apj] {10.1086/171708}, \href
  {https://ui.adsabs.harvard.edu/abs/1992ApJ...396..178F} {396, 178}

\bibitem[\protect\citeauthoryear{{Fryxell} et~al.,}{{Fryxell}
  et~al.}{2000}]{Fryxell2000}
{Fryxell} B.,  et~al., 2000, \mn@doi [\apjs] {10.1086/317361}, \href
  {https://ui.adsabs.harvard.edu/abs/2000ApJS..131..273F} {131, 273}

\bibitem[\protect\citeauthoryear{{Gavagnin}, {Bleuler}, {Rosdahl}  \&
  {Teyssier}}{{Gavagnin} et~al.}{2017}]{Gavagnin2017}
{Gavagnin} E.,  {Bleuler} A.,  {Rosdahl} J.,   {Teyssier} R.,  2017, \mn@doi
  [\mnras] {10.1093/mnras/stx2222}, \href
  {https://ui.adsabs.harvard.edu/abs/2017MNRAS.472.4155G} {472, 4155}

\bibitem[\protect\citeauthoryear{{Gehrels}}{{Gehrels}}{1986}]{Gehrels1986}
{Gehrels} N.,  1986, \mn@doi [\apj] {10.1086/164079}, \href
  {https://ui.adsabs.harvard.edu/abs/1986ApJ...303..336G} {303, 336}

\bibitem[\protect\citeauthoryear{{Gorti} \& {Bhatt}}{{Gorti} \&
  {Bhatt}}{1996}]{Gorti1996}
{Gorti} U.,  {Bhatt} H.~C.,  1996, \mn@doi [\mnras] {10.1093/mnras/283.2.566},
  \href {https://ui.adsabs.harvard.edu/abs/1996MNRAS.283..566G} {283, 566}

\bibitem[\protect\citeauthoryear{{Heggie}}{{Heggie}}{1975}]{Heggie1975}
{Heggie} D.~C.,  1975, \mn@doi [\mnras] {10.1093/mnras/173.3.729}, \href
  {https://ui.adsabs.harvard.edu/abs/1975MNRAS.173..729H} {173, 729}

\bibitem[\protect\citeauthoryear{{Heggie} \& {Rasio}}{{Heggie} \&
  {Rasio}}{1996}]{Heggie1996}
{Heggie} D.~C.,  {Rasio} F.~A.,  1996, \mn@doi [\mnras]
  {10.1093/mnras/282.3.1064}, \href
  {https://ui.adsabs.harvard.edu/abs/1996MNRAS.282.1064H} {282, 1064}

\bibitem[\protect\citeauthoryear{{Hills}}{{Hills}}{1975}]{Hills1975}
{Hills} J.~G.,  1975, \mn@doi [\aj] {10.1086/111815}, \href
  {https://ui.adsabs.harvard.edu/abs/1975AJ.....80..809H} {80, 809}

\bibitem[\protect\citeauthoryear{Hughes \& Hase}{Hughes \&
  Hase}{2010}]{measurements}
Hughes I.~G.,  Hase T.~P.,  2010, Measurements and their Uncertainties.
Oxford University Press, Oxford

\bibitem[\protect\citeauthoryear{{Hut}, {Makino}  \& {McMillan}}{{Hut}
  et~al.}{1995}]{Hut1995}
{Hut} P.,  {Makino} J.,   {McMillan} S.,  1995, \mn@doi [\apjl]
  {10.1086/187844}, \href
  {https://ui.adsabs.harvard.edu/abs/1995ApJ...443L..93H} {443, L93}

\bibitem[\protect\citeauthoryear{{Ivanova}, {Heinke}, {Rasio}, {Taam},
  {Belczynski}  \& {Fregeau}}{{Ivanova} et~al.}{2006}]{Ivanova2006}
{Ivanova} N.,  {Heinke} C.~O.,  {Rasio} F.~A.,  {Taam} R.~E.,  {Belczynski} K.,
    {Fregeau} J.,  2006, \mn@doi [\mnras] {10.1111/j.1365-2966.2006.10876.x},
  \href {https://ui.adsabs.harvard.edu/abs/2006MNRAS.372.1043I} {372, 1043}

\bibitem[\protect\citeauthoryear{{King}, {Goodwin}, {Parker}  \&
  {Patience}}{{King} et~al.}{2012}]{King2012}
{King} R.~R.,  {Goodwin} S.~P.,  {Parker} R.~J.,   {Patience} J.,  2012,
  \mn@doi [\mnras] {10.1111/j.1365-2966.2012.22108.x}, \href
  {https://ui.adsabs.harvard.edu/abs/2012MNRAS.427.2636K} {427, 2636}

\bibitem[\protect\citeauthoryear{{Kobulnicky} et~al.,}{{Kobulnicky}
  et~al.}{2014}]{Kobulnicky2014}
{Kobulnicky} H.~A.,  et~al., 2014, \mn@doi [\apjs]
  {10.1088/0067-0049/213/2/34}, \href
  {https://ui.adsabs.harvard.edu/abs/2014ApJS..213...34K} {213, 34}

\bibitem[\protect\citeauthoryear{{Kouwenhoven}, {Brown}, {Zinnecker}, {Kaper}
  \& {Portegies Zwart}}{{Kouwenhoven} et~al.}{2005}]{Kouwenhoven2005}
{Kouwenhoven} M.~B.~N.,  {Brown} A.~G.~A.,  {Zinnecker} H.,  {Kaper} L.,
  {Portegies Zwart} S.~F.,  2005, \mn@doi [\aap] {10.1051/0004-6361:20048124},
  \href {https://ui.adsabs.harvard.edu/abs/2005A&A...430..137K} {430, 137}

\bibitem[\protect\citeauthoryear{{Kouwenhoven}, {Brown}, {Goodwin}, {Portegies
  Zwart}  \& {Kaper}}{{Kouwenhoven} et~al.}{2009}]{Kouwenhoven2009}
{Kouwenhoven} M.~B.~N.,  {Brown} A.~G.~A.,  {Goodwin} S.~P.,  {Portegies Zwart}
  S.~F.,   {Kaper} L.,  2009, \mn@doi [\aap] {10.1051/0004-6361:200810234},
  \href {https://ui.adsabs.harvard.edu/abs/2009A&A...493..979K} {493, 979}

\bibitem[\protect\citeauthoryear{{Kouwenhoven}, {Goodwin}, {Parker}, {Davies},
  {Malmberg}  \& {Kroupa}}{{Kouwenhoven} et~al.}{2010}]{Kouwenhoven2010}
{Kouwenhoven} M.~B.~N.,  {Goodwin} S.~P.,  {Parker} R.~J.,  {Davies} M.~B.,
  {Malmberg} D.,   {Kroupa} P.,  2010, \mn@doi [\mnras]
  {10.1111/j.1365-2966.2010.16399.x}, \href
  {https://ui.adsabs.harvard.edu/abs/2010MNRAS.404.1835K} {404, 1835}

\bibitem[\protect\citeauthoryear{{Kroupa}}{{Kroupa}}{1995}]{Kroupa1995}
{Kroupa} P.,  1995, \mn@doi [\mnras] {10.1093/mnras/277.4.1522}, \href
  {https://ui.adsabs.harvard.edu/abs/1995MNRAS.277.1522K} {277, 1522}

\bibitem[\protect\citeauthoryear{{Kroupa}}{{Kroupa}}{2001}]{kroupa}
{Kroupa} P.,  2001, \mn@doi [\mnras] {10.1046/j.1365-8711.2001.04022.x}, \href
  {https://ui.adsabs.harvard.edu/abs/2001MNRAS.322..231K} {322, 231}

\bibitem[\protect\citeauthoryear{{Kruijssen}, {Pelupessy}, {Lamers}, {Portegies
  Zwart}  \& {Icke}}{{Kruijssen} et~al.}{2011}]{Kruijssen2011}
{Kruijssen} J.~M.~D.,  {Pelupessy} F.~I.,  {Lamers} H. J.~G.~L.~M.,  {Portegies
  Zwart} S.~F.,   {Icke} V.,  2011, \mn@doi [\mnras]
  {10.1111/j.1365-2966.2011.18467.x}, \href
  {https://ui.adsabs.harvard.edu/abs/2011MNRAS.414.1339K} {414, 1339}

\bibitem[\protect\citeauthoryear{{Krumholz}, {McKee}  \&
  {Bland-Hawthorn}}{{Krumholz} et~al.}{2019}]{Krumholz2019}
{Krumholz} M.~R.,  {McKee} C.~F.,   {Bland-Hawthorn} J.,  2019, \mn@doi [\araa]
  {10.1146/annurev-astro-091918-104430}, \href
  {https://ui.adsabs.harvard.edu/abs/2019ARA&A..57..227K} {57, 227}

\bibitem[\protect\citeauthoryear{{Lada} \& {Lada}}{{Lada} \&
  {Lada}}{2003}]{Lada2003}
{Lada} C.~J.,  {Lada} E.~A.,  2003, \mn@doi [\araa]
  {10.1146/annurev.astro.41.011802.094844}, \href
  {https://ui.adsabs.harvard.edu/abs/2003ARA&A..41...57L} {41, 57}

\bibitem[\protect\citeauthoryear{{Lee}}{{Lee}}{2013}]{Lee2013}
{Lee} D.,  2013, \mn@doi [J. Comput. Phys.] {10.1016/j.jcp.2013.02.049}, \href
  {https://ui.adsabs.harvard.edu/abs/2013JCoPh.243..269L} {243, 269}

\bibitem[\protect\citeauthoryear{{Lee}, {Lee}, {Dunham}, {Tatematsu}, {Choi},
  {Bergin}  \& {Evans}}{{Lee} et~al.}{2017}]{Lee2017}
{Lee} J.-E.,  {Lee} S.,  {Dunham} M.~M.,  {Tatematsu} K.,  {Choi} M.,  {Bergin}
  E.~A.,   {Evans} N.~J.,  2017, \mn@doi [Nature Astronomy]
  {10.1038/s41550-017-0172}, \href
  {https://ui.adsabs.harvard.edu/abs/2017NatAs...1E.172L} {1, 0172}

\bibitem[\protect\citeauthoryear{{Leigh}, {Giersz}, {Webb}, {Hypki},
  {\VAN{Marchi}{De}{de} Marchi}, {Kroupa}  \& {Sills}}{{Leigh}
  et~al.}{2013}]{Leigh2013}
{Leigh} N.,  {Giersz} M.,  {Webb} J.~J.,  {Hypki} A.,  {\VAN{Marchi}{De}{de}
  Marchi} G.,  {Kroupa} P.,   {Sills} A.,  2013, \mn@doi [\mnras]
  {10.1093/mnras/stt1825}, \href
  {https://ui.adsabs.harvard.edu/abs/2013MNRAS.436.3399L} {436, 3399}

\bibitem[\protect\citeauthoryear{Mann \& Whitney}{Mann \&
  Whitney}{1947}]{Mann1947}
Mann H.~B.,  Whitney D.~R.,  1947, \mn@doi [Ann. Math. Statist.]
  {10.1214/aoms/1177730491}, 18, 50

\bibitem[\protect\citeauthoryear{Mardling}{Mardling}{2008}]{Mardling2008}
Mardling R.~A.,  2008, in Vesperini E.,  Giersz M.,   Sills A.,  eds,
  Proceedings IAU Symposium No. 246. Cambridge University Press, Cambridge, pp
  199--208

\bibitem[\protect\citeauthoryear{{Matzner} \& {McKee}}{{Matzner} \&
  {McKee}}{2000}]{Matzner2000}
{Matzner} C.~D.,  {McKee} C.~F.,  2000, \mn@doi [\apj] {10.1086/317785}, \href
  {https://ui.adsabs.harvard.edu/abs/2000ApJ...545..364M} {545, 364}

\bibitem[\protect\citeauthoryear{{McMillan} \& {Hut}}{{McMillan} \&
  {Hut}}{1996}]{McMillan1996}
{McMillan} S. L.~W.,  {Hut} P.,  1996, \mn@doi [\apj] {10.1086/177610}, \href
  {https://ui.adsabs.harvard.edu/abs/1996ApJ...467..348M} {467, 348}

\bibitem[\protect\citeauthoryear{{McMillan}, {Portegies Zwart}, {van Elteren}
  \& {Whitehead}}{{McMillan} et~al.}{2012}]{McMillan2012}
{McMillan} S.,  {Portegies Zwart} S.,  {van Elteren} A.,   {Whitehead} A.,
  2012, in {Capuzzo-Dolcetta} R.,  {Limongi} M.,   {Tornamb{\`e}} A.,  eds,
  ASP Conference Series Vol. 453, Advances in Computational Astrophysics:
  Methods, Tools, and Outcome. San Francisco, p.~129 (\mn@eprint {arXiv}
  {1111.3987})

\bibitem[\protect\citeauthoryear{{Miholics}, {Kruijssen}  \&
  {Sills}}{{Miholics} et~al.}{2017}]{Miholics2017}
{Miholics} M.,  {Kruijssen} J.~M.~D.,   {Sills} A.,  2017, \mn@doi [\mnras]
  {10.1093/mnras/stx1312}, \href
  {https://ui.adsabs.harvard.edu/abs/2017MNRAS.470.1421M} {470, 1421}

\bibitem[\protect\citeauthoryear{{Milone} et~al.,}{{Milone}
  et~al.}{2016}]{Milone2016}
{Milone} A.~P.,  et~al., 2016, \mn@doi [\mnras] {10.1093/mnras/stv2415}, \href
  {https://ui.adsabs.harvard.edu/abs/2016MNRAS.455.3009M} {455, 3009}

\bibitem[\protect\citeauthoryear{{Miyoshi} \& {Kusano}}{{Miyoshi} \&
  {Kusano}}{2005}]{Miyoshi2005}
{Miyoshi} T.,  {Kusano} K.,  2005, \mn@doi [J. Comput. Phys.]
  {10.1016/j.jcp.2005.02.017}, \href
  {https://ui.adsabs.harvard.edu/abs/2005JCoPh.208..315M} {208, 315}

\bibitem[\protect\citeauthoryear{{Moe} \& {{\VAN{Stefano}{Di}{di}}
  Stefano}}{{Moe} \& {{\VAN{Stefano}{Di}{di}} Stefano}}{2015}]{Moe2015}
{Moe} M.,  {{\VAN{Stefano}{Di}{di}} Stefano} R.,  2015, \mn@doi [\apj]
  {10.1088/0004-637X/810/1/61}, \href
  {https://ui.adsabs.harvard.edu/abs/2015ApJ...810...61M} {810, 61}

\bibitem[\protect\citeauthoryear{{Moe} \& {{\VAN{Stefano}{Di}{di}}
  Stefano}}{{Moe} \& {{\VAN{Stefano}{Di}{di}} Stefano}}{2017}]{Moe2017}
{Moe} M.,  {{\VAN{Stefano}{Di}{di}} Stefano} R.,  2017, \mn@doi [\apjs]
  {10.3847/1538-4365/aa6fb6}, \href
  {https://ui.adsabs.harvard.edu/abs/2017ApJS..230...15M} {230, 15}

\bibitem[\protect\citeauthoryear{{\VAN{Neumann}{Von}{von}
  Neumann}}{{\VAN{Neumann}{Von}{von} Neumann}}{1951}]{Neumann1951}
{\VAN{Neumann}{Von}{von} Neumann} J.,  1951, in Householder A.~S.,  Forsythe
  G.~E.,   Germond H.~H.,  eds, National Bureau of Standards Applied
  Mathematics Series, Vol.~12, Monte Carlo Method.
US Government Printing Office, Washington, DC, Chapt.~13, pp 36--38

\bibitem[\protect\citeauthoryear{{Offner}, {Kratter}, {Matzner}, {Krumholz}  \&
  {Klein}}{{Offner} et~al.}{2010}]{offner}
{Offner} S. S.~R.,  {Kratter} K.~M.,  {Matzner} C.~D.,  {Krumholz} M.~R.,
  {Klein} R.~I.,  2010, \mn@doi [\apj] {10.1088/0004-637X/725/2/1485}, \href
  {https://ui.adsabs.harvard.edu/abs/2010ApJ...725.1485O} {725, 1485}

\bibitem[\protect\citeauthoryear{Parker \& Goodwin}{Parker \&
  Goodwin}{2012}]{Parker2012}
Parker R.~J.,  Goodwin S.~P.,  2012, \mn@doi [MNRAS]
  {10.1111/j.1365-2966.2012.21190.x}, 424, 272

\bibitem[\protect\citeauthoryear{{Parker} \& {Meyer}}{{Parker} \&
  {Meyer}}{2014}]{Parker2014}
{Parker} R.~J.,  {Meyer} M.~R.,  2014, \mn@doi [\mnras]
  {10.1093/mnras/stu1101}, \href
  {https://ui.adsabs.harvard.edu/abs/2014MNRAS.442.3722P} {442, 3722}

\bibitem[\protect\citeauthoryear{{Parker} \& {Reggiani}}{{Parker} \&
  {Reggiani}}{2013}]{Parker2013}
{Parker} R.~J.,  {Reggiani} M.~M.,  2013, \mn@doi [\mnras]
  {10.1093/mnras/stt600}, \href
  {https://ui.adsabs.harvard.edu/abs/2013MNRAS.432.2378P} {432, 2378}

\bibitem[\protect\citeauthoryear{{Parker}, {Goodwin}, {Kroupa}  \&
  {Kouwenhoven}}{{Parker} et~al.}{2009}]{parker}
{Parker} R.~J.,  {Goodwin} S.~P.,  {Kroupa} P.,   {Kouwenhoven} M.~B.~N.,
  2009, \mn@doi [\mnras] {10.1111/j.1365-2966.2009.15032.x}, \href
  {https://ui.adsabs.harvard.edu/abs/2009MNRAS.397.1577P} {397, 1577}

\bibitem[\protect\citeauthoryear{{Pelupessy} \& {Portegies Zwart}}{{Pelupessy}
  \& {Portegies Zwart}}{2012}]{Pelupessy2012}
{Pelupessy} F.~I.,  {Portegies Zwart} S.,  2012, \mn@doi [\mnras]
  {10.1111/j.1365-2966.2011.20137.x}, \href
  {https://ui.adsabs.harvard.edu/abs/2012MNRAS.420.1503P} {420, 1503}

\bibitem[\protect\citeauthoryear{Portegies~Zwart \& McMillan}{Portegies~Zwart
  \& McMillan}{2019}]{amuse}
Portegies~Zwart S.,  McMillan S.~L.~W.,  2019, Astrophysical Recipes: The Art
  of Amuse.
Institute of Physics Publishing, Bristol

\bibitem[\protect\citeauthoryear{{Portegies Zwart} \& {Verbunt}}{{Portegies
  Zwart} \& {Verbunt}}{1996}]{seba}
{Portegies Zwart} S.~F.,  {Verbunt} F.,  1996, \aap, \href
  {https://ui.adsabs.harvard.edu/abs/1996A&A...309..179P} {309, 179}

\bibitem[\protect\citeauthoryear{{Portegies Zwart}, {Makino}, {McMillan}  \&
  {Hut}}{{Portegies Zwart} et~al.}{1999}]{PortegiesZwart1999}
{Portegies Zwart} S.~F.,  {Makino} J.,  {McMillan} S.~L.~W.,   {Hut} P.,  1999,
  \aap, \href {https://ui.adsabs.harvard.edu/abs/1999A&A...348..117P} {348,
  117}

\bibitem[\protect\citeauthoryear{{Portegies Zwart}, {McMillan}, {Hut}  \&
  {Makino}}{{Portegies Zwart} et~al.}{2001}]{PortegiesZwart2001}
{Portegies Zwart} S.~F.,  {McMillan} S. L.~W.,  {Hut} P.,   {Makino} J.,  2001,
  \mn@doi [\mnras] {10.1046/j.1365-8711.2001.03976.x}, \href
  {https://ui.adsabs.harvard.edu/abs/2001MNRAS.321..199P} {321, 199}

\bibitem[\protect\citeauthoryear{{Portegies Zwart}, {McMillan}  \&
  {Gieles}}{{Portegies Zwart} et~al.}{2010}]{PortegiesZwart2010}
{Portegies Zwart} S.~F.,  {McMillan} S. L.~W.,   {Gieles} M.,  2010, \mn@doi
  [\araa] {10.1146/annurev-astro-081309-130834}, \href
  {https://ui.adsabs.harvard.edu/abs/2010ARA&A..48..431P} {48, 431}

\bibitem[\protect\citeauthoryear{Press, Teukolsky, Vetterling  \&
  Flannery}{Press et~al.}{2007}]{recipes}
Press W.~H.,  Teukolsky S.~A.,  Vetterling W.~T.,   Flannery B.~P.,  2007,
  Numerical Recipes: The Art of Scientific Computing, 3rd edn.
Cambridge University Press, Cambridge

\bibitem[\protect\citeauthoryear{{Price} \& {Bate}}{{Price} \&
  {Bate}}{2008}]{Price2008}
{Price} D.~J.,  {Bate} M.~R.,  2008, \mn@doi [\mnras]
  {10.1111/j.1365-2966.2008.12976.x}, \href
  {https://ui.adsabs.harvard.edu/abs/2008MNRAS.385.1820P} {385, 1820}

\bibitem[\protect\citeauthoryear{Price-Whelan et~al.,}{Price-Whelan
  et~al.}{2020}]{Price-Whelan2020}
Price-Whelan A.~M.,  et~al., 2020, \mn@doi [\apj] {10.3847/1538-4357/ab8acc},
  \href {https://ui.adsabs.harvard.edu/abs/2020ApJ...895....2P} {895, 2}

\bibitem[\protect\citeauthoryear{{Raghavan} et~al.,}{{Raghavan}
  et~al.}{2010}]{Raghavan2010}
{Raghavan} D.,  et~al., 2010, \mn@doi [\apjs] {10.1088/0067-0049/190/1/1},
  \href {https://ui.adsabs.harvard.edu/abs/2010ApJS..190....1R} {190, 1}

\bibitem[\protect\citeauthoryear{{Rastello}, {Carraro}  \&
  {Capuzzo-Dolcetta}}{{Rastello} et~al.}{2020}]{Rastello2020}
{Rastello} S.,  {Carraro} G.,   {Capuzzo-Dolcetta} R.,  2020, \mn@doi [\apj]
  {10.3847/1538-4357/ab910b}, \href
  {https://ui.adsabs.harvard.edu/abs/2020ApJ...896..152R} {896, 152}

\bibitem[\protect\citeauthoryear{Reid \& Gizis}{Reid \& Gizis}{1997}]{Reid1997}
Reid I.~N.,  Gizis J.~E.,  1997, AJ, 113, 2246

\bibitem[\protect\citeauthoryear{{Reipurth}, {Guimar{\~a}es}, {Connelley}  \&
  {Bally}}{{Reipurth} et~al.}{2007}]{Reipurth2007}
{Reipurth} B.,  {Guimar{\~a}es} M.~M.,  {Connelley} M.~S.,   {Bally} J.,  2007,
  \mn@doi [\aj] {10.1086/523596}, \href
  {https://ui.adsabs.harvard.edu/abs/2007AJ....134.2272R} {134, 2272}

\bibitem[\protect\citeauthoryear{Reipurth, Clarke, Boss, Goodwin, Rodríguez,
  Stassun, Tokovinin  \& Zinnecker}{Reipurth et~al.}{2014}]{Reipurth2014}
Reipurth B.,  Clarke C.~J.,  Boss A.~P.,  Goodwin S.~P.,  Rodríguez L.~F.,
  Stassun K.~G.,  Tokovinin A.,   Zinnecker H.,  2014, in Beuther H.,  {et al.}
  eds, , Protostars and Planets VI.
University of Arizona, Tucson, pp 267--290

\bibitem[\protect\citeauthoryear{{Ricker}}{{Ricker}}{2008}]{Ricker2008}
{Ricker} P.~M.,  2008, \mn@doi [\apjs] {10.1086/526425}, \href
  {https://ui.adsabs.harvard.edu/abs/2008ApJS..176..293R} {176, 293}

\bibitem[\protect\citeauthoryear{{\VAN{Rosa}{De}{de} Rosa}
  et~al.,}{{\VAN{Rosa}{De}{de} Rosa} et~al.}{2014}]{DeRosa2014}
{\VAN{Rosa}{De}{de} Rosa} R.~J.,  et~al., 2014, \mn@doi [\mnras]
  {10.1093/mnras/stt1932}, \href
  {https://ui.adsabs.harvard.edu/abs/2014MNRAS.437.1216D} {437, 1216}

\bibitem[\protect\citeauthoryear{{Sana} \& {Evans}}{{Sana} \&
  {Evans}}{2011}]{Sana2011}
{Sana} H.,  {Evans} C.~J.,  2011, in {Neiner} C.,  {Wade} G.,  {Meynet} G.,
  {Peters} G.,  eds,  IAU Symposium Vol. 272, Active OB Stars: Structure,
  Evolution, Mass Loss, and Critical Limits. Cambridge University Press,
  Cambridge, pp 474--485

\bibitem[\protect\citeauthoryear{{Sana} et~al.,}{{Sana}
  et~al.}{2012}]{Sana2012}
{Sana} H.,  et~al., 2012, \mn@doi [Science] {10.1126/science.1223344}, \href
  {https://ui.adsabs.harvard.edu/abs/2012Sci...337..444S} {337, 444}

\bibitem[\protect\citeauthoryear{{Sigalotti}, {Cruz}, {Gabbasov}, {Klapp}  \&
  {Ram{\'\i}rez-Velasquez}}{{Sigalotti} et~al.}{2018}]{Sigalotti2018}
{Sigalotti} L. D.~G.,  {Cruz} F.,  {Gabbasov} R.,  {Klapp} J.,
  {Ram{\'\i}rez-Velasquez} J.,  2018, \mn@doi [\apj]
  {10.3847/1538-4357/aab619}, \href
  {https://ui.adsabs.harvard.edu/abs/2018ApJ...857...40S} {857, 40}

\bibitem[\protect\citeauthoryear{{Sigurdsson} \& {Phinney}}{{Sigurdsson} \&
  {Phinney}}{1993}]{1993ApJ...415..631S}
{Sigurdsson} S.,  {Phinney} E.~S.,  1993, \mn@doi [\apj] {10.1086/173190},
  \href {https://ui.adsabs.harvard.edu/abs/1993ApJ...415..631S} {415, 631}

\bibitem[\protect\citeauthoryear{Sills \& Bailyn}{Sills \&
  Bailyn}{1999}]{Sills1999}
Sills A.,  Bailyn C.~D.,  1999, ApJ, 513, 428

\bibitem[\protect\citeauthoryear{Sills, Rieder, Scora, McCloskey  \&
  Jaffa}{Sills et~al.}{2018}]{Sills2018}
Sills A.,  Rieder S.,  Scora J.,  McCloskey J.,   Jaffa S.,  2018, MNRAS, 477,
  1903

\bibitem[\protect\citeauthoryear{{Sollima}}{{Sollima}}{2008}]{Sollima2008}
{Sollima} A.,  2008, \mn@doi [\mnras] {10.1111/j.1365-2966.2008.13387.x}, \href
  {https://ui.adsabs.harvard.edu/abs/2008MNRAS.388..307S} {388, 307}

\bibitem[\protect\citeauthoryear{{Sormani}, {Tre{\ss}}, {Klessen}  \&
  {Glover}}{{Sormani} et~al.}{2017}]{Sormani2017}
{Sormani} M.~C.,  {Tre{\ss}} R.~G.,  {Klessen} R.~S.,   {Glover} S. C.~O.,
  2017, \mn@doi [\mnras] {10.1093/mnras/stw3205}, \href
  {https://ui.adsabs.harvard.edu/abs/2017MNRAS.466..407S} {466, 407}

\bibitem[\protect\citeauthoryear{{Stahler}}{{Stahler}}{2010}]{Stahler2010}
{Stahler} S.~W.,  2010, \mn@doi [\mnras] {10.1111/j.1365-2966.2009.15994.x},
  \href {https://ui.adsabs.harvard.edu/abs/2010MNRAS.402.1758S} {402, 1758}

\bibitem[\protect\citeauthoryear{{Tobin} et~al.,}{{Tobin}
  et~al.}{2016a}]{tobin2016b}
{Tobin} J.~J.,  et~al., 2016a, \mn@doi [\nat] {10.1038/nature20094}, \href
  {https://ui.adsabs.harvard.edu/abs/2016Natur.538..483T} {538, 483}

\bibitem[\protect\citeauthoryear{{Tobin} et~al.,}{{Tobin}
  et~al.}{2016b}]{Tobin2016a}
{Tobin} J.~J.,  et~al., 2016b, \mn@doi [\apj] {10.3847/0004-637X/818/1/73},
  \href {https://ui.adsabs.harvard.edu/abs/2016ApJ...818...73T} {818, 73}

\bibitem[\protect\citeauthoryear{{Tokovinin}}{{Tokovinin}}{2017}]{tokovinin}
{Tokovinin} A.,  2017, \mn@doi [\mnras] {10.1093/mnras/stx707}, \href
  {https://ui.adsabs.harvard.edu/abs/2017MNRAS.468.3461T} {468, 3461}

\bibitem[\protect\citeauthoryear{{Wall}, {McMillan}, {Mac Low}, {Klessen}  \&
  {Portegies Zwart}}{{Wall} et~al.}{2019}]{wall1}
{Wall} J.~E.,  {McMillan} S. L.~W.,  {Mac Low} M.-M.,  {Klessen} R.~S.,
  {Portegies Zwart} S.,  2019, \mn@doi [\apj] {10.3847/1538-4357/ab4db1}, \href
  {https://ui.adsabs.harvard.edu/abs/2019ApJ...887...62W} {887, 62}

\bibitem[\protect\citeauthoryear{Wall, Mac~Low, McMillan, Klessen,
  Portegies~Zwart  \& Pellegrino}{Wall et~al.}{2020}]{wall2}
Wall J.~E.,  Mac~Low M.-M.,  McMillan S.~L.~W.,  Klessen R.~S.,
  Portegies~Zwart S.,   Pellegrino A.,  2020, submitted to ApJ

\bibitem[\protect\citeauthoryear{Wilcoxon}{Wilcoxon}{1945}]{Wilcoxon1945}
Wilcoxon F.,  1945, \mn@doi [Biometrics Bulletin] {10.2307/3001968}, 1, 80

\bibitem[\protect\citeauthoryear{{Winters} et~al.,}{{Winters}
  et~al.}{2019}]{winters}
{Winters} J.~G.,  et~al., 2019, \mn@doi [\aj] {10.3847/1538-3881/ab05dc}, \href
  {https://ui.adsabs.harvard.edu/abs/2019AJ....157..216W} {157, 216}

\makeatother
\end{thebibliography}

%%%%%%%%%%%%%%%%%%%%%%%%%%%%%%%%%%%%%%%%%%%%%%%%%%

%%%%%%%%%%%%%%%%% APPENDICES %%%%%%%%%%%%%%%%%%%%%

%\appendix

%%%%%%%%%%%%%%%%%%%%%%%%%%%%%%%%%%%%%%%%%%%%%%%%%%

% Don't change these lines
\bsp	% typesetting comment
\label{lastpage}
\end{document}